\documentclass[twocolumn,showpacs,preprintnumbers,amsmath,amssymb]{revtex4}

\usepackage{graphicx}
\usepackage{dcolumn}
\usepackage{bm}

\begin{document}

\preprint{}

\title{Efficient Heating of Thin Cylindrical Targets\\by Broad Electromagnetic Beams I}

\author{Andrey Akhmeteli}
 \email{akhmeteli@home.domonet.ru}
\affiliation{%
38-2-139 Rublevskoe shosse\\
121609 Moscow, Russia
}%


\date{May 18, 2004}

\begin{abstract}
In many high-profile applications, such as nuclear fusion and pumping of active media of short-wavelength lasers, it is necessary to achieve high specific input of power of an electromagnetic beam in a target. Diffraction sets the lower limit to the transverse dimensions of electromagnetic beams and represents a fundamental obstacle for electromagnetic heating of small or inaccessible regions.
It was found, however, that it is possible to achieve efficient heating of cylindrical targets by electromagnetic beams with transverse dimensions that are several orders of magnitude greater than those of the cylinder. These counter-intuitive results have the following physical mechanism: the absorption in the cylinder causes a deep fall in the field distribution, and this fall causes diffractive diffusion of the field towards the axis from a large volume of the beam.
The heating efficiency was rigorously calculated using the exact solution of the problem of diffraction on an infinite homogeneous cylinder (J.R. Wait, 1955). Several non-resonant domains of parameters were found that provide efficient absorption of the energy of a broad beam in a thin conducting cylinder. The typical asymptotic efficiency for very thin cylinders is (2-5)/L, where L=ln(w/d), where d is the diameter of the cylinder and w is the width of the beam waist in the longitudinal geometry or the wavelength in the transverse geometry. This efficiency is high even if the beam is several orders of magnitude broader than the cylinder.
The relevant conditions are not too rigid and may be used to heat cylindrical targets to high temperatures in a number of important applications. External magnetic field may further relax the conditions of efficient heating.
\end{abstract}

\maketitle

\section{\label{sec:level11}Introduction}

In many high-profile applications it is necessary to achieve high specific input of power of an electromagnetic beam in a target. No-contact methods are of special interest, as the resulting temperatures may be hardly compatible with any walls. Heating by electromagnetic waves is the  most important of these methods. It is used for controlled fusion, for pumping of active media of short-wavelength lasers, in plasma chemical reactors, for material processing, and for many other applications. Diffraction sets lower limits to the transverse dimensions of the electromagnetic beam and thus presents a fundamental obstacle for electromagnetic heating of small or distant regions. One may use shorter wavelengths to circumvent the diffraction constraints, but this is usually connected with very serious technical difficulties and lower efficiency. Another possible way out is multiple-pass heating of small conducting regions -- e.g., in resonators or waveguides. But in this case heating is unacceptably slow for many applications, and it is practically impossible to heat distant regions (e.g., to form an ionized structure in the atmosphere for broadcast or radar). Non-linear effects, such as self-focusing, may be very important, but they typically require an electron-density minimum in the region under heating and/or high beam intensity.

\begin{figure}
\includegraphics{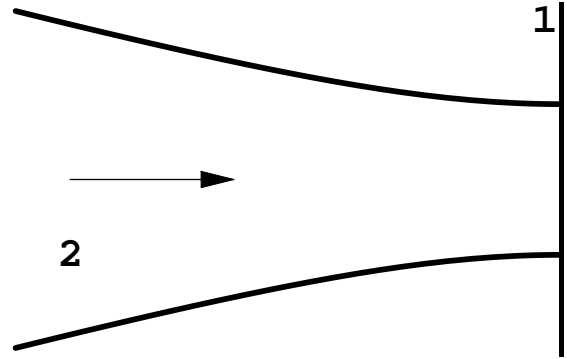}
\caption{\label{fig:gr2} Standard geometry.
1 -- flat target; 2 -- incident Gaussian electromagnetic beam.}
\end{figure}

\begin{figure}
\includegraphics{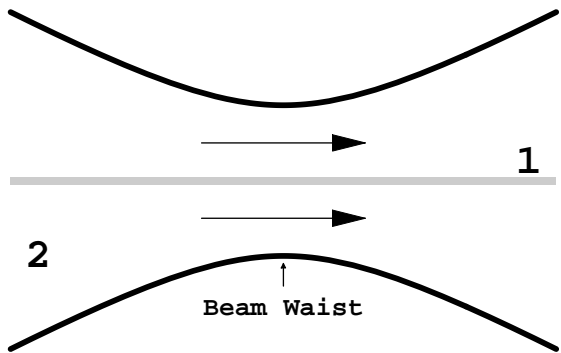}
\caption{\label{fig:gr1} Longitudinal geometry (the axes of a Gaussian beam and a cylinder coincide; the energy flows mainly along the axis).
1 -- cylindrical target;
2 -- incident Gaussian electromagnetic beam (energy flows from left to right).}
\end{figure}

\begin{figure}
\includegraphics{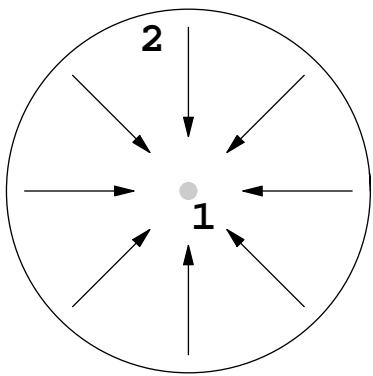}
\caption{\label{fig:gr4} Transverse geometry (the axes of a converging cylindrical wave and a cylinder coincide; there is no energy flow along the axis; the wavelength is considered as "the transverse dimension of the beam"). 1 -- cylindrical target;
2 –- incident converging cylindrical electromagnetic wave .}
\end{figure}

Thus, in standard geometries of target irradiation, only one dimension of the region under heating may be much smaller than the transverse dimensions of the electromagnetic beam (see Fig.~\ref{fig:gr2}).
On the other hand, it was found that in the longitudinal geometry (Fig.~\ref{fig:gr1}) and the transverse geometry (Fig.~\ref{fig:gr4}), if certain, not too rigid, conditions are met, it is possible to achieve efficient fast heating of cylindrical targets by electromagnetic beams with transverse dimensions that are several orders of magnitude greater than those of the cylinder (Ref.~\cite{Akhm1}, \cite{Akhm2}, \cite{Akhm3}, \cite{Akhm4}, \cite{Akhm5}, \cite{Akhm6}, \cite{Akhm7}, \cite{Akhm8}, \cite{Akhm9}). These counter-intuitive results have the following physical mechanism: the absorption in the cylinder causes a deep fall in the field distribution, and this fall causes diffractive diffusion of the field towards the axis from a large volume of the beam. The effect is linear (for given properties of the cylinder, the heating efficiency does not depend on the beam power) and scalable (may find applications in various frequency ranges).
As these results are rather surprising, much effort was made to reliably confirm them. For instance, for the longitudinal geometry, three methods were used: a qualitative ("back-of-the envelope") estimate, a semiquantitative ("one-wave") calculation, and a quantitative calculation, where the Maxwell equations were solved virtually without any approximations. The latter was used to confirm the results of the semiquantitative approach, which was used to derive an asymptotic formula for heating efficiency and determine the parameter domains that provide high efficiency.

In the qualitative estimate (Section~\ref{sec:level14}), it is assumed that the values of the frequency of the electromagnetic field  and the conductivity of the cylinder are such that the skin-depth and the diameter of the cylinder are of the same order of magnitude. Then it is assumed that in this case the refracted electric field inside the cylinder has the same magnitude as the electric field in the incident beam ("the electromagnetic field penetrates the cylinder to the skin-depth"). Then we obtain that the Joule heating power is of the same order of magnitude as the power in the incident beam.

Of course, diffraction problems are notorious for strong dependence on the parameters, and this primitive estimate is often incorrect, but it gives reasonable order-of-magnitude results in some important cases and demonstrates the scaling of the problem. For example, if we reduce the diameter of the cylinder diameter by half, the optimal conductivity increases fourfold (as the skin-depth is inversely proportional to the square root of the conductivity), and the absorbed power remains roughly the same. Alternatively, if the width of the beam waist increases twofold, the electric field in the cylinder falls foufold, but the length of the beam waist increases fourfold, so the absorption efficiency does not change.

In the semiquantitative approach, it is assumed that the refracted electric field inside the overlap of the cylinder and of the Gaussian beam waist is the same as the refracted electric field for a cylindrical wave that has the field intensity on the axis equal to that in the center of the Gaussian beam and the width of the first maximum equal to the width of the beam waist. Then the exact solution of the problem of diffraction of a cylindrical wave on an infinite homogeneous cylinder is used (Ref.~\cite{Wait1}).

The quantitative calculation was performed as follows:

1. For rigorous calculation of diffraction of a Gaussian beam on an infinite homogeneous cylinder, exact solutions of the free Maxwell equations were found (Section~\ref{sec:level12}) that are approximated by Gaussian beams with great accuracy when the beam waist radius is much greater than the wavelength (if this condition is not fulfilled, a Gaussian beam is not a good approximate solution of the free Maxwell equations). This result may be important for many applications.

2. For the cylindrical waves, the Wait's solution was calculated (in this exact solution of the Maxwell equations, each cylindrical wave incident on an infinite homogeneous cylinder generates two reflected and two refracted cylindrical waves).

3. The resulting refracted fields were calculated by integration of the fields of the refracted cylindrical waves using fast Hankel transform.

4. The heating efficiency was calculated as a ratio of the power of the Joule heating in the cylinder (within the length of the waist of the Gaussian beam) to the power in the incident free Gaussian beam.

Thus, virtually no approximations were used in this rigorous calculation. The results of the semiquantitative approach coincided with those of the rigorous calculation with 50\% accuracy.

Let us describe the structure of this first part of the work.

In Section~\ref{sec:level12}, exact solutions of the free Maxwell equations are described that are approximated by Gaussian beams with great accuracy when the beam waist radius is much greater than the wavelength, and their asymptotic properties are proved. It is also proven that these solutions are normalizable.

In Section~\ref{sec:level13}, the exact solution of the problem of diffraction of a cylindrical wave on an infinite homogeneous cylinder (Ref.~\cite{Wait1}) is summarized.

In Section~\ref{sec:level14}, the qualitative and the semiquantitative approaches are described. In the framework of the latter, an asymptotic formula for heating efficiency in the longitudinal geometry is derived.

In Section~\ref{sec:level15}, an asymptotic formula for heating efficiency in the transverse geometry is derived. This calculation is relatively simple and easy to verify, but has many common features with the more complex calculations for the longitudinal geometry.

In Section~\ref{sec:level16}, some conclusions are formulated.

Thus, this first part contains considerable technical material. In the second part, the parameter domains are determined that provide efficient heating in the longitudinal geometry. The relevant conditions were formulated in part in Ref.~\cite{Akhm6}, \cite{Akhm7}; they refine and generalize the results of Ref.~\cite{Akhm1}. The conditions are confirmed by rigorous calculation, and some applications are discussed in the second part.
\maketitle
\section{\label{sec:level12}Exact solutions of the free Maxwell equations and Gaussian beams}
It is highly desirable to validate the results of the present paper through comparison with exact solutions of the Maxwell equations. However, Gaussian beams are just approximations to exact solutions of free Maxwell equations. So in this section the author informally derives some exact solutions of the free Maxwell equations and proves that Gaussian beams approximate them with high accuracy. These solutions were found in Ref.~\cite{Akhm8}. In Ref.~\cite{Lekner1}, drawbacks of the earlier approaches to this problem are discussed, and other beam-like solutions of the free Maxwell equations are found.

We consider a monochromatic beam with frequency $\omega$ and use such a system of units that the wave vector in free space $k_{0}=\frac{\omega}{c}=1$. Solutions of the Maxwell equations in homogeneous media with permittivity $\varepsilon$ and magnetic permeability $\mu$ are described by the electric and magnetic Hertz vectors \bm{$\Pi$} and \bm{$\Pi'$}. The Hertz vectors satisfy equations:
\begin{equation}\label{eq:5}
 \triangle\bm{\Pi}+k^{2}\bm{\Pi}=0,
\end{equation}
\begin{equation}\label{eq:6}
 \triangle\bm{\Pi'}+k^{2}\bm{\Pi'}=0,
\end{equation}
where $k^{2}=\frac{\varepsilon\mu\omega^{2}}{c^{2}}$.
The electric and magnetic fields \bm{$E$} and \bm{$H$} can be calculated from the Hertz vectors as follows:
\begin{equation}\label{eq:3}
\bm{H}=-ik_{0}\bm{\nabla}\bm{\times}\bm{\Pi}+\frac{1}{\mu}\bm{\nabla}\bm{\times}(\bm{\nabla}\bm{\times}\bm{\Pi'}),
\end{equation}
\begin{equation}\label{eq:4}
\bm{E}=\frac{1}{\varepsilon}\bm{\nabla}\bm{\times}(\bm{\nabla}\bm{\times}\bm{\Pi})+ik_{0}\bm{\nabla}\bm{\times}\bm{\Pi'}.
\end{equation}
In the cylindrical coordinates $\{\rho,\varphi,z\}$ or in the cartesian coordinates $\{x,y,z\}$ we can choose the Hertz vectors in the following form:
 \begin{equation}\label{eq:1}
    \bm{\Pi}=\{0,0,\Pi\},
\end{equation}
\begin{equation}\label{eq:2}
    \bm{\Pi'}=\{0,0,\Pi'\}.
\end{equation}
In free space $\varepsilon=\mu=1$.
To emulate a Gaussian electromagnetic beam with circular polarization let us assume that in the symmetry plane of the beam ($z=0$) we have
\begin{equation}\label{eq:7}
    \bm{E}(x,y,0)=\{i,-1,0\}\exp(-\frac{x^{2}+y^{2}}{2\delta^{2}}),
\end{equation}
\begin{equation}\label{eq:8}
    \bm{H}(x,y,0)=\{1,i,0\}\exp(-\frac{x^{2}+y^{2}}{2\delta^{2}}).
\end{equation}
It should be noted that, for the sake of simplicity, some conventions are used here, so, the actual electric and magnetic fields are equal to the real parts of the products of the expressions in  Eqs.~(\ref{eq:7}) and ~(\ref{eq:8}) by the time-dependent factor $\exp(-i\omega t)$.
Let us write the Fourier expansions of Eqs.~(\ref{eq:7}) and ~(\ref{eq:8}) :
\begin{eqnarray}\label{eq:9}
\bm{E}(x,y,0)=&&\{i,-1,0\}\frac{\delta^{2}}{2\pi}
\int\exp\left(-\delta^{2}\frac{k_{x}^{2}+k_{y}^{2}}{2}\right)\nonumber\\
&&\times
\exp(i(k_{x}x+k_{y}y))\,dk_{x}dk_{y},
\end{eqnarray}
\begin{eqnarray}\label{eq:10}
\bm{H}(x,y,0)=&&\{1,i,0\}\frac{\delta^{2}}{2\pi}
\int\exp\left(-\delta^{2}\frac{k_{x}^{2}+k_{y}^{2}}{2}\right)\nonumber\\
&&\times
\exp(i(k_{x}x+k_{y}y))\,dk_{x}dk_{y},
\end{eqnarray}
However, these expressions are not compatible with the free Maxwell equations, as, e.g., on the one hand, $H_{z}(x,y,0)\equiv 0$, therefore $\dot{H}_{z}(x,y,0)\equiv0$ and $(\bm{\nabla}\bm{\times}\bm{E})_{z}(x,y,0)\equiv 0$, and the latter is not true. So we have to modify these expressions. The Fourier expansions suggest that the fields may be constructed as a superposition of plane waves.
Let us consider a plane electromagnetic wave with a wave vector $\{0,0,1\}$  -- it propagates in the direction $z$. To obtain a plane wave with a wave vector ${k_{x},k_{y},k_{z}}$, where $k_{x}^{2}+k_{y}^{2}+k_{z}=1$, we may multiply the field vectors by the matrix
\begin{eqnarray}\label{eq:9a}
\bm{A}=&&\left(%
\begin{array}{ccc}
  k_{z} & 0 & k_{x} \\
  0 & k_{z} & k_{y} \\
  -k_{x} & -k_{y} & k_{z} \\
\end{array}%
\right)+\nonumber\\
&&
+\frac{1}{1+k_{z}}\left(%
\begin{array}{ccc}
  k_{y}^{2} & -k_{x}k_{y} & 0 \\
  -k_{x}k_{y} & k_{x}^{2}& 0 \\
  0 & 0 & 0 \\
\end{array}%
\right)
\end{eqnarray}
and change the phase factor. Following such reasoning, we obtain
(or, rather, choose)
\begin{eqnarray}\label{eq:12}
E_{z}(x,y,0)=&&\
\frac{ w_0^{2}}{2\pi}\int\exp\left(- w_0^{2}\frac{k_{x}^{2}+k_{y}^{2}}{2}\right)(-i k_{x}+k_{y})
\nonumber\\
&&\times
\exp(i(k_{x}x+k_{y}y))\,dk_{x}dk_{y},
\end{eqnarray}
\begin{equation}\label{eq:13}
    H_{z}(x,y,0)=-i E_{z}(x,y,0).
\end{equation}
As
$E_{z}=-\triangle_{2}\Pi=-(\partial^{2}_{x}+\partial^{2}_{y})\Pi$,
\begin{eqnarray}\label{eq:14}
\Pi(x,y,0)=&&\
\frac{ w_0^{2}}{2\pi}\int\exp\left(- w_0^{2}\frac{k_{x}^{2}+k_{y}^{2}}{2}\right)\frac{-i k_{x}+k_{y}}{k_{x}^{2}+k_{y}^{2}}
\nonumber\\
&&\times
\exp(i(k_{x}x+k_{y}y))\,dk_{x}dk_{y}=\nonumber\\
=&&\
\frac{ w_0^{2}}{2\pi}\int\exp\left(- w_0^{2}\frac{k_{x}^{2}+k_{y}^{2}}{2}\right)\frac{1}{i k_{x}+k_{y}}
\nonumber\\
&&\times
\exp(i(k_{x}x+k_{y}y))\,dk_{x}dk_{y}.
\end{eqnarray}
Let us change the variables: $k_{x}=\rho\cos \varphi$, $k_{y}=\rho\sin \varphi$:
\begin{eqnarray}\label{eq:15}
\Pi(x,y,0)=&&\
\frac{ w_0^{2}}{2\pi}\int\exp\left(- w_0^{2}\frac{\rho^{2}}{2}\right)\frac{\rho}{\rho(\sin \varphi+i\cos \varphi)}
\nonumber\\
&&\times
\exp(i\rho(x\cos\varphi+y\sin \varphi))d\rho d\varphi=\nonumber\\
=&&\
\frac{ w_0^{2}}{2\pi}\int\exp\left(- w_0^{2}\frac{\rho^{2}}{2}\right)\frac{1}{i\exp(-i\varphi)}
\nonumber\\
&&\times
\exp(i\rho(x\cos\varphi+y\sin \varphi))d\rho d\varphi.
\end{eqnarray}
In the cylindrical coordinates ($x=r\cos\alpha$, $y=r\sin\alpha$), as $\cos\alpha\cos\varphi+\sin\alpha\sin \varphi=\cos(\alpha-\varphi)$, we have:
\begin{eqnarray}\label{eq:16}
\Pi(r,\alpha,0)=&&\
\frac{ w_0^{2}}{2\pi}\int_0^\infty\exp\left(- w_0^{2}\frac{\rho^{2}}{2}\right)
\int_0^{2\pi}(-i\exp(i\varphi))
\nonumber\\
&&\times
\exp(i\rho r\cos(\alpha-\varphi))\,d\rho d\varphi.
\end{eqnarray}
Let us calculate the following integral:
\begin{eqnarray}\label{eq:17}
\int_0^{2\pi}\exp(i\rho r\cos(\alpha-\varphi))(-i\exp(i\varphi))d\varphi=
\nonumber\\
=-i\int_0^{2\pi}\exp(i\rho r\cos(\alpha-\varphi))\exp(i(\varphi-\alpha))
\exp(i\alpha)d\varphi=
\nonumber\\
=-i\exp(i\alpha)\int_{-\alpha}^{2\pi-\alpha}\exp(i\rho r\cos\varphi)
\exp(i\varphi)d\varphi=
\nonumber\\
=-i\exp(i\alpha)\int_{0}^{2\pi}\exp(i\rho r\cos\varphi)\exp(i\varphi)d\varphi=
\nonumber\\
=-i\exp(i\alpha)2\pi i J_{1}(\rho r)=2\pi\exp(i\alpha)J_{1}(\rho r)\qquad
\end{eqnarray}
($J_{k}(x)$ is the Bessel function).
Thus,
\begin{equation}\label{eq:18}
\Pi(r,\alpha,0)= w_0^{2}\exp(i\alpha)
\int_0^\infty\exp\left(- w_0^{2}\frac{\rho^{2}}{2}\right)J_{1}(\rho r)d\rho,
\end{equation}
or, if we change the variables $\rho\rightarrow\lambda$, $r\rightarrow\rho$, $\alpha\rightarrow\varphi$,
\begin{equation}\label{eq:19}
\Pi(\rho,\varphi,0)= w_0^{2}\exp(i\varphi)
\int_0^\infty\exp\left(- w_0^{2}\frac{\lambda^{2}}{2}\right)J_{1}(\lambda\rho)\,d\lambda.
\end{equation}
It can be shown that
\begin{equation}\label{eq:20}
\Pi(\rho,\varphi,0)=\frac{w_0^2}{\rho}\exp(i\varphi)
\left(1-\exp\left(-\frac{\rho^{2}}{2 w_0^{2}}\right)\right).
\end{equation}

We need to know the fields for all values of $z$. We choose the following expressions for $z$-components of the Hertz vectors:
\begin{eqnarray}\label{eq:21}
\Pi(x,y,z)=&&\
\frac{ w_0^{2}}{2\pi}\int\limits_{k_{x}^{2}+k_{y}^{2}\leq 1}\exp\left(- w_0^{2}\frac{k_{x}^{2}+k_{y}^{2}}{2}\right)\frac{1}{i k_{x}+k_{y}}
\nonumber\\
&&\times
\exp\left(i\left(k_{x}x+k_{y}y+\sqrt{1-k_{x}^{2}-k_{y}^{2}}\,z\right)\right)
\nonumber\\
&&\times\,dk_{x}dk_{y},
\nonumber\\
\Pi'=-i\Pi.
\end{eqnarray}
We have limited the region of integration by the condition $k_{x}^{2}+k_{y}^{2}\leq 1$ to eliminate reactive fields and leave running waves only.
In Eq.~(\ref{eq:21}) the solution is presented as a linear combination of plane waves. For our purpose it is preferable to present the solution as a linear combination of cylindrical waves:
\begin{eqnarray}\label{eq:22}
\Pi(\rho,\varphi,z)= w_0^{2}\exp(i\varphi)
\int_0^1\exp\left(- w_0^{2}\frac{\lambda^{2}}{2}\right)
\nonumber\\
\times\exp\left(i\sqrt{1-\lambda^2}\,z\right)J_{1}(\lambda\rho)\,d\lambda,
\nonumber\\
\Pi'=-i\Pi.
\end{eqnarray}
Let us find the asymptotics of Eq.~(\ref{eq:22}) for the case $ w_0\gg1$ (when the Gaussian beam waist width is much greater than the wavelength).
Let us assume that $\alpha$ is a large parameter . In view of the scaling of Gaussian beams (the beam waist length increases as the square of its width), we assume that $ w_0=\alpha w_0'$, $\rho=\alpha\rho'$, $z=\alpha^2 z'$, where $ w_0'$, $\rho'$, and $z'$ are fixed, and $\alpha\rightarrow\infty$.
\begin{eqnarray}\label{eq:23}
\Pi(\rho,\varphi,z)=\alpha^2 w_0'^{2}\exp(i\varphi)
\int_0^1\exp\left(- w_0'^{2}\alpha^2\frac{\lambda^{2}}{2}\right)
\nonumber\\
\times\exp\left(i\sqrt{1-\lambda^2}\,\alpha^2 z'\right)J_{1}(\lambda\alpha\rho')\,d\lambda.
\end{eqnarray}
\begin{eqnarray}\label{eq:24}
\sqrt{1-\lambda^2}=1-\frac{\lambda^2}{2}-\left(1-\frac{\lambda^2}{2}-\sqrt{1-\lambda^2}\right)=
\nonumber\\
=1-\frac{\lambda^2}{2}-\frac{\left(1-\frac{\lambda^2}{2}\right)^2-\left(\sqrt{1-\lambda^2}\,\right)^2}
{1-\frac{\lambda^2}{2}+\sqrt{1-\lambda^2}}=
\nonumber\\
=1-\frac{\lambda^2}{2}-\frac{\lambda^4}
{4\left(1-\frac{\lambda^2}{2}+\sqrt{1-\lambda^2}\,\right)},
\end{eqnarray}
therefore,
\begin{eqnarray}\label{eq:25}
\Pi(\rho,\varphi,z)=\alpha^2 w_0'^{2}\exp(i\varphi)
\int_0^1\exp\left(- w_0'^{2}\alpha^2\frac{\lambda^{2}}{2}\right)
\nonumber\\
\times\exp\left(i\left(1-\frac{\lambda^2}{2}\right)\alpha^2 z'\right)
\nonumber\\
\times\exp\left(-i\frac{\lambda^4}
{4\left(1-\frac{\lambda^2}{2}+\sqrt{1-\lambda^2}\,\right)}\,\alpha^2 z'\right)J_{1}(\lambda\alpha\rho')\,d\lambda.
\end{eqnarray}
Let us change the integration variable: $x=\lambda\alpha\rho'$.
\begin{eqnarray}\label{eq:26}
\Pi(\rho,\varphi,z)=\alpha^2 w_0'^{2}\exp(i\varphi)
\nonumber\\
\times
\int_0^{\alpha\rho'}\exp(i\alpha^2 z')\exp\left(-\frac{x^2 w_0'^{2}}{2\rho'^2}\right)
\exp\left(-i\frac{x^2 z'}{2\rho'^2}\right)
\nonumber\\
\times\exp\left(-i\frac{x^4\alpha^{-2}\rho'^{-4} z'}
{4\left(1-\frac{x^2}{2\alpha^2\rho'^2}+\sqrt{1-\frac{x^2}{\alpha^2\rho'^2}}\,\right)}\right)\,J_{1}(x)
\frac{dx}{\alpha\rho'}=
\nonumber\\
=\frac{\alpha w_0'^{2}}{\rho'}\exp(i\varphi)\exp(i\alpha^2 z')
\nonumber\\
\times\int_0^{\alpha\rho'}\exp\left(-\frac{x^2}{2\rho'^2}( w_0'^2+i z')\right)
\nonumber\\
\times\exp\left(-i\frac{x^4\alpha^{-2}\rho'^{-4} z'}
{4\left(1-\frac{x^2}{2\alpha^2\rho'^2}+\sqrt{1-\frac{x^2}{\alpha^2\rho'^2}}\,\right)}\right)\,J_{1}(x)
\,dx.\quad
\end{eqnarray}
Let us assume that $\rho'\neq 0$ and $\alpha^{1/4}<\alpha\rho'$ ($\alpha$ is great enough). Let us also note that
\begin{equation}\label{eq:27}
1/2\leq f(\lambda)=1-\frac{\lambda^2}{2}+\sqrt{1-\lambda^2}\,\leq 2,
\end{equation}
on the segment $0\leq\lambda\leq1$, as $f(\lambda)$ is monotone decreasing on this segment, and its values at the ends of the segment are $2$ and $1/2$. We also note that $\vert J_{1}(x)\vert<1$ for real $x$.

The last integral in Eq.~(\ref{eq:26}) may be presented in the following form:
\begin{equation}\label{eq:28}
\int_0^{\alpha\rho'}=\int_0^{\alpha^{1/4}}+\int_{\alpha^{1/4}}^{\alpha\rho'}.
\end{equation}
We have:
\begin{eqnarray}\label{eq:29}
\left\vert\int_{\alpha^{1/4}}^{\alpha\rho'}\right\vert\leq
\nonumber\\
\leq\left\vert\exp\left(-\frac{\alpha^{1/2}}{2\rho'^2}( w_0'^2+i z')\right)\right\vert\left\vert\alpha\rho'-\alpha^{1/4}\right\vert
\rightarrow 0
\end{eqnarray}
for $\alpha\rightarrow\infty$.
\begin{eqnarray}\label{eq:30}
\int_0^{\alpha^{1/4}}=\int_0^{\alpha^{1/4}}\exp\left(-\frac{x^2}{2\rho'^2}( w_0'^2+i z')\right)
\nonumber\\
\times(1+\rm{O}(\alpha^{-1}))J_{1}(x)
\,dx,
\end{eqnarray}
\begin{eqnarray}\label{eq:31}
\left\vert\int_0^{\alpha^{1/4}}\exp\left(-\frac{x^2}{2\rho'^2}( w_0'^2+i z')\right)
\times \rm{O}(\alpha^{-1})J_{1}(x)\,dx\right\vert\leq
\nonumber\\
\leq\left\vert \rm{O}(\alpha^{-1})\right\vert\alpha^{1/4}\rightarrow 0\quad
\end{eqnarray}
for $\alpha\rightarrow\infty$.
\begin{eqnarray}\label{eq:32}
\int_0^{\alpha^{1/4}}\exp\left(-\frac{x^2}{2\rho'^2}( w_0'^2+i z')\right)
\times J_{1}(x)\,dx=\int_0^{\infty}+\rm{o}(1),
\nonumber\\
\int_0^{\infty}\exp\left(-\frac{x^2}{2\rho'^2( w_0'^2+i z')}\right)J_{1}(x)\,dx=
\nonumber\\
=1-\exp\left(-\frac{\rho'^2}{2( w_0'^2+i z')}\right)\quad
\end{eqnarray}
(cf. Ref.~\cite{Prudn1}, p.186, formula 2.12.9.2).
Therefore, for large $\alpha$
\begin{eqnarray}\label{eq:33}
\Pi(\rho,\varphi,z)\approx\frac{\alpha w_0'^2}{\rho'}\exp(i\varphi)\exp(i\alpha^2 z')
\nonumber\\
\times\left(1-\exp\left(-\frac{\rho'^2}{2( w_0'^2+i z')}\right)\right)=
\nonumber\\
=\frac{ w_0^2}{\rho}\exp(i\varphi)\exp(i z)\left(1-\exp\left(-\frac{\rho^2}{2( w_0^2+i z)}\right)\right).
\end{eqnarray}
We also have
\begin{equation}\label{eq:34}
\Pi'=-i\Pi.
\end{equation}
In free space $\varepsilon=1$ and $\mu=1$, therefore,
\begin{equation}\label{eq:35}
\bm{H}=-i\bm{\nabla}\bm{\times}\bm{\Pi}-i\bm{\nabla}\bm{\times}(\bm{\nabla}\bm{\times}\bm{\Pi}),
\end{equation}
\begin{equation}\label{eq:36}
\bm{E}=\bm{\nabla}\bm{\times}(\bm{\nabla}\bm{\times}\bm{\Pi})+\bm{\nabla}\bm{\times}\bm{\Pi},
\end{equation}
thus,
\begin{equation}\label{eq:37}
\bm{E}=i\bm{H}.
\end{equation}
Taking into consideration Eq.~(\ref{eq:1}), we have in the cylindrical coordinates:
\begin{equation}\label{eq:38}
\bm{\nabla}\bm{\times}\bm{\Pi}=\left\{\frac{1}{\rho}\frac{\partial\Pi}{\partial\varphi},
-\frac{\partial\Pi}{\partial\rho},0\right\},
\end{equation}
\begin{widetext}
\begin{eqnarray}\label{eq:39}
\bm{\nabla}\bm{\times}(\bm{\nabla}\bm{\times}\bm{\Pi})=
\left\{-\frac{\partial}{\partial z}\left(-\frac{\partial\Pi}{\partial\rho}\right),
\frac{\partial}{\partial z}\left(\frac{1}{\rho}\frac{\partial\Pi}{\partial\varphi}\right),\frac{1}{\rho}
\left(\frac{\partial}{\partial\rho}\left(\rho\left(-\frac{\partial\Pi}{\partial\rho}\right)\right)
-\frac{\partial}{\partial\varphi}\left(\frac{1}{\rho}
\frac{\partial\Pi}{\partial\varphi}\right)\right)\right\},
\end{eqnarray}
\end{widetext}
\begin{equation}\label{eq:40}
\frac{\partial\Pi}{\partial\varphi}=i\Pi
\end{equation}
Let us present $\Pi$ in the following form:
\begin{equation}\label{eq:41}
\Pi=\Pi_1+\Pi_2,
\end{equation}
where
\begin{equation}\label{eq:42}
\Pi_1=\frac{ w_0^2}{\rho}\exp(i\varphi)\exp(i z),
\end{equation}
\begin{equation}\label{eq:43}
\Pi_2=-\frac{ w_0^2}{\rho}\exp(i\varphi)\exp(i z)exp\left(-\frac{\rho^2}{2( w_0^2+i z)}\right).
\end{equation}
We have:
\begin{equation}\label{eq:44}
\frac{\partial\Pi_1}{\partial\varphi}=i\Pi_1,
\end{equation}
\begin{equation}\label{eq:45}
\frac{\partial\Pi_1}{\partial z}=i\Pi_1,
\end{equation}
\begin{equation}\label{eq:46}
\frac{\partial\Pi_1}{\partial\rho}=-\frac{1}{\rho}\Pi_1,
\end{equation}
\begin{equation}\label{eq:47}
\bm{\nabla}\bm{\times}\bm{\Pi_1}=\left\{\frac{i}{\rho}\Pi_1,
\frac{1}{\rho}\Pi_1,0\right\},
\end{equation}
\begin{widetext}
\begin{eqnarray}\label{eq:48}
\bm{\nabla}\bm{\times}(\bm{\nabla}\bm{\times}\bm{\Pi_1})=
\left\{-\frac{\partial}{\partial z}\left(\frac{1}{\rho}\Pi_1\right),
\frac{\partial}{\partial z}\left(\frac{i}{\rho}\Pi_1\right),\frac{1}{\rho}
\left(\frac{\partial}{\partial\rho}\left(Pi_1\right)
-\frac{\partial}{\partial\varphi}\left(\frac{i}{\rho}
\Pi_1\right)\right)\right\}=
\nonumber\\
=\left\{-\frac{i}{\rho}\Pi_1,-\frac{1}{\rho}\Pi_1,
\frac{1}{\rho}\left(-\frac{1}{\rho}\Pi_1+\frac{1}{\rho}\Pi_1\right)\right\},
\end{eqnarray}
\end{widetext}
therefore, $\Pi_1$ does not contribute to the fields.
\begin{equation}\label{eq:49}
\frac{\partial\Pi_2}{\partial\varphi}=i\Pi_2,
\end{equation}
\begin{eqnarray}\label{eq:50}
\frac{\partial\Pi_2}{\partial z}=i\Pi_2+\Pi_2\frac{\rho^2}{2( w_0^2+i z)^2}i=
\nonumber\\
=i\Pi_2\left(1+\frac{\rho^2}{2( w_0^2+i z)^2}\right),
\end{eqnarray}
\begin{eqnarray}\label{eq:51}
\frac{\partial\Pi_2}{\partial\rho}=\Pi_2\frac{-\rho}{ w_0^2+i z}-\frac{1}{\rho}\Pi_2=\Pi_2\left(\frac{-\rho}{ w_0^2+i z}-\frac{1}{\rho}\right)=
\nonumber\\
=-\Pi_2\left(\frac{\rho}{ w_0^2+i z}+\frac{1}{\rho}\right),
\end{eqnarray}
\begin{eqnarray}\label{eq:52}
\bm{\nabla}\bm{\times}\bm{\Pi_2}=\left\{\frac{i}{\rho}\Pi_2,\left(\frac{\rho}{ w_0^2+i z}+\frac{1}{\rho}\right)\Pi_2,0\right\},
\end{eqnarray}
\begin{widetext}
\begin{eqnarray}\label{eq:53}
(\bm{\nabla}\bm{\times}(\bm{\nabla}\bm{\times}\bm{\Pi_2}))_{\rho}=\frac{\partial}{\partial z}\left(-\Pi_2\left(\frac{\rho}{ w_0^2+i z}+\frac{1}{\rho}\right)\right)=
-i\Pi_2\left(1+\frac{\rho^2}{2( w_0^2+i z)^2}\right)\left(\frac{\rho}{ w_0^2+i z}+\frac{1}{\rho}\right)-\Pi_2\frac{-\rho}{( w_0^2+i z)^2}i=
\nonumber\\
=-i\Pi_2\left(\frac{\rho}{ w_0^2+i z}+\frac{1}{\rho}+\frac{\rho^3}{2( w_0^2+i z)^3}
+\frac{\rho}{2( w_0^2+i z)^2}-\frac{\rho}{( w_0^2+i z)^2}\right),\quad
\end{eqnarray}
\begin{eqnarray}\label{eq:54}
(\bm{\nabla}\bm{\times}(\bm{\nabla}\bm{\times}\bm{\Pi_2}))_{\varphi}=\frac{\partial}{\partial z}\left(\frac{i}{\rho}\Pi_2\right)=\frac{i}{\rho}i\Pi_2\left(1+\frac{\rho^2}{2( w_0^2+i z)^2}\right)=-\Pi_2\left(\frac{1}{\rho}+\frac{\rho}{2( w_0^2+i z)^2}\right),
\end{eqnarray}
\begin{eqnarray}\label{eq:55}
(\bm{\nabla}\bm{\times}(\bm{\nabla}\bm{\times}\bm{\Pi_2}))_z=\frac{1}{\rho}\left(\frac{\partial}{\partial\rho}
\left(\Pi_2\left(\frac{\rho^2}{ w_0^2+i z}+1\right)\right)+\frac{1}{\rho}\Pi_2\right)=
\nonumber\\
=-\Pi_2\left(\frac{1}{\rho}+\frac{\rho}{2( w_0^2+i z)^2}\right)=
\frac{1}{\rho}\left(-\Pi_2\left(\frac{\rho}{ w_0^2+i z}+\frac{1}{\rho}\right)
\left(\frac{\rho^2}{ w_0^2+i z}+1\right)+\Pi_2\frac{2\rho}{ w_0^2+i z}+\frac{1}{\rho} \Pi_2\right)=
\nonumber\\
=\Pi_2\left(-\frac{\rho^2}{( w_0^2+i z)^2}-\frac{1}{ w_0^2+i z}-\frac{1}{ w_0^2+i z}-\frac{1}{\rho^2}+\frac{2}{ w_0^2+i z}+\frac{1}{\rho^2}\right)=-\Pi_2\frac{\rho^2}{( w_0^2+i z)^2},
\end{eqnarray}
\begin{eqnarray}\label{eq:56}
\bm{\nabla}\bm{\times}\bm{\Pi_2}+\bm{\nabla}\bm{\times}(\bm{\nabla}\bm{\times}\bm{\Pi_2})=
\nonumber\\
=
\left\{-i\Pi_2\left(\frac{\rho}{ w_0^2+i z}+\frac{\rho^3}{2( w_0^2+i z)^3}+\frac{\rho}{2( w_0^2+i z)^2}-\frac{\rho}{( w_0^2+i z)^2}\right),\Pi_2\left(\frac{\rho}{ w_0^2+i z}-\frac{\rho}{2( w_0^2+i z)^2}\right),\Pi_2\left(-\frac{\rho^2}{( w_0^2+i z)^2}\right)\right\}=
\nonumber\\
=- w_0^2\exp(i\varphi)\exp(i z)\exp\left(-\frac{\rho^2}{2( w_0^2+i z)}\right)\frac{1}{ w_0^2+i z}\left\{-i\left(1+\frac{\rho^2}{2( w_0^2+i z)^2}-\frac{1}{2( w_0^2+i z)}\right),1-\frac{1}{2( w_0^2+i z)},-\frac{\rho}{ w_0^2+i z}\right\}.\quad
\end{eqnarray}
\end{widetext}
If $\varphi=0$, $z=0$, $\rho=0$, $ w_0\gg 1$, the fields have the following components:
\begin{equation}\label{eq:57}
\bm{E}=\bm{\nabla}\bm{\times}\bm{\Pi_2}+\bm{\nabla}\bm{\times}(\bm{\nabla}\bm{\times}\bm{\Pi_2})=
\{i,-1,0\},
\end{equation}
\begin{equation}\label{eq:58}
\bm{H}=-i\bm{E}=\{1,i,0\}
\end{equation}
With an accuracy of $\rm{o}(1)$ the field components may be presented in the following form:
\begin{eqnarray}\label{eq:59}
\bm{E}=\frac{ w_0^2}{ w_0^2+i z}\exp(i\varphi)\exp(i z)
\nonumber\\
\times\exp\left(-\frac{\rho^2}{2( w_0^2+i z)}\right)\{i,-1,0\},
\end{eqnarray}
\begin{eqnarray}\label{eq:60}
\bm{H}=\frac{ w_0^2}{ w_0^2+i z}\exp(i\varphi)\exp(i z)
\nonumber\\
\times\exp\left(-\frac{\rho^2}{2( w_0^2+i z)}\right)\{1,i,0\},
\end{eqnarray}
(The field components are not small in comparison with $1$ only if $\rho\lesssim\sqrt{\vert w_0^2+i z\vert}$; we also take into account that $ w_0\gg 1$).
Thus, the fields have the space distribution typical for a Gaussian beam. However, so far the fields were calculated from the Hertz vectors of Eqs.~(\ref{eq:33}) and (\ref{eq:34}), so let us prove that the same asymptotic expressions are valid for the Hertz vectors of Eq.~(\ref{eq:22}).
From Eqs.~(\ref{eq:22}) and (\ref{eq:24}) we have:
\begin{equation}\label{eq:61}
\Pi(\rho,\varphi,z)= w_0^2\exp(i\varphi)\int_0^1 f(\lambda)J_1(\lambda\rho)\,d\lambda,
\end{equation}
where
\begin{eqnarray}\label{eq:62}
f(\lambda)=\exp\left(-\frac{\lambda^2 w_0^2}{2}\right)
\nonumber\\
\times\exp\left(i z\left(1-\frac{\lambda^2}{2}-\frac{\lambda^4}
{4\left(1-\frac{\lambda^2}{2}+\sqrt{1-\lambda^2}\,\right)}\right)\right).
\end{eqnarray}
We need to calculate the derivatives of $\Pi(\rho,\varphi,z)$:
\begin{equation}\label{eq:63}
\frac{\partial\Pi}{\partial\varphi}=i\Pi,
\end{equation}
and to obtain $\frac{\partial\Pi}{\partial z}$, $\frac{\partial\Pi}{\partial\rho}$, and $\frac{\partial^2\Pi}{\partial\rho^2}$ we need to make the following replacements in the right-hand side of Eq.~(\ref{eq:61}):
\begin{eqnarray}\label{eq:64}
f(\lambda)\rightarrow i \left(1-\frac{\lambda^2}{2}-\frac{\lambda^4}
{4\left(1-\frac{\lambda^2}{2}+\sqrt{1-\lambda^2}\,\right)}\right)f(\lambda),
\nonumber\\
J_1(\lambda\rho)\rightarrow\lambda J'_1(\lambda\rho),
\nonumber\\
J_1(\lambda\rho)\rightarrow\lambda^2 J''_1(\lambda\rho),
\end{eqnarray}
correspondingly. We have:
\begin{widetext}
\begin{eqnarray}\label{eq:65}
\bm{E}=\bm{\nabla}\bm{\times}\bm{\Pi}+\bm{\nabla}\bm{\times}(\bm{\nabla}\bm{\times}\bm{\Pi})=
\left\{\frac{1}{\rho}\frac{\partial\Pi}{\partial\varphi}+\frac{\partial^2\Pi}{\partial z\partial\rho},\frac{1}{\rho}\frac{\partial^2\Pi}{\partial z\partial\varphi}-\frac{\partial\Pi}{\partial\rho},\frac{1}
{\rho}\left(-\frac{\partial\Pi}{\partial\rho}-\rho\frac{\partial^2\Pi}{\partial\rho^2}-
\frac{1}{\rho}\frac{\partial^2\Pi}{\partial\varphi^2}\right)\right\}=
\nonumber\\
=\left\{\frac{i}{\rho}\Pi+\frac{\partial^2\Pi}{\partial z\partial\rho},\frac{i}{\rho}\frac{\partial\Pi}{\partial z}-\frac{\partial\Pi}{\partial\rho},-\frac{1}
{\rho}\frac{\partial\Pi}{\partial\rho}-\frac{\partial^2\Pi}{\partial\rho^2}+
\frac{1}{\rho^2}\Pi\right\}.
\end{eqnarray}
\end{widetext}
The terms with $\Pi$ in the last line of Eq.~(\ref{eq:65}) may be estimated in the same way as $\Pi$ in Eq.~(\ref{eq:23}) and further. For the terms with $\frac{\partial\Pi}{\partial\rho}$ and $\frac{\partial^2\Pi}{\partial\rho^2}$ we should make the following replacement:
\begin{eqnarray}\label{eq:66}
J_1(\lambda\rho)\rightarrow\left\{0,-\lambda J'_1(\lambda\rho),-\frac{\lambda}{\rho} J'_1(\lambda\rho)-\lambda^2 J''_1((\lambda\rho)\right\}.\quad
\end{eqnarray}
In Eq.~(\ref{eq:26}) we make the replacement:
\begin{eqnarray}\label{eq:67}
J_1(x)\rightarrow\left\{0,-\frac{x}{\alpha\rho'}J'_1(x),-\frac{x}{\alpha\rho'} J'_1(x)-\frac{x^2}{\alpha^2\rho'^2} J''_1(x)\right\}.\quad
\end{eqnarray}
The estimates closely follow those in Eq.~(\ref{eq:26}) and further. The remaining integral analogous to that in Eq.~(\ref{eq:32}) may be calculated by differentiation of the latter with respect to a parameter.
For the terms with $\frac{\partial^2\Pi}{\partial z\partial\rho}$ and $\frac{\partial\Pi}{\partial z}$ in Eq.~(\ref{eq:65}) we make the replacement of Eq.~(\ref{eq:64}) for $f(\lambda)$ and the following replacement:
\begin{eqnarray}\label{eq:68}
J_1(\lambda\rho)\rightarrow\left\{\lambda J'_1(\lambda\rho),\frac{i}{\rho} J_1(\lambda\rho),0\right\}.\quad
\end{eqnarray}
To estimate these terms we make the following replacements in Eq.~(\ref{eq:26}): the factor
\begin{eqnarray}\label{eq:69}
g=i\left(1-\frac{x^2}{2\alpha^2\rho'^2}-\frac{x^4\alpha^{-4}\rho^{-4}}
{4\left(1-\frac{x^2}{2\alpha^2\rho'^2}+\sqrt{1-\frac{x^2}{\alpha^2\rho'^2}}\,\right)}\right)
\end{eqnarray}
is added, and
\begin{eqnarray}\label{eq:70}
J_1(x)\rightarrow\left\{\frac{x}{\alpha\rho'}J'_1(x),-\frac{i}{\alpha\rho'}J_1(x),0\right\}.\quad
\end{eqnarray}
Then we follow the estimates of Eq.~(\ref{eq:26}) and further. The remaining integral has the following form:
\begin{eqnarray}\label{eq:71}
\int_0^{\alpha^{1/4}}\exp\left(-\frac{x^2}{2\rho'^2}( w_0^2+i z')\right)g
\nonumber\\
\times\left\{\frac{x}{\alpha\rho'}J'_1(x),-\frac{i}{\alpha\rho'}J_1(x),0\right\}\,dx.\quad
\end{eqnarray}
The first two terms in the expression for $g$ (Eq.~(\ref{eq:69})) give the contributions that are estimated in the same way as the integral in Eq.~(\ref{eq:32}) and present its derivatives.
For the remaining integral
\begin{widetext}
\begin{eqnarray}\label{eq:72}
I=\int_0^{\alpha^{1/4}}\exp\left(-\frac{x^2}{2\rho'^2}( w_0^2+i z')\right)\left(-\frac{x^4\alpha^{-4}\rho^{-4}}
{4\left(1-\frac{x^2}{2\alpha^2\rho'^2}+\sqrt{1-\frac{x^2}{\alpha^2\rho'^2}}\,\right)}\right)
\left\{\frac{x}{\alpha\rho'}J'_1(x),\frac{i}{\alpha\rho'}J_1(x),0\right\}\,dx\quad
\end{eqnarray}
\end{widetext}
we have $\vert I\vert\leq\alpha^{-3}\rm{o}(1)$, so it does not make significant contributions to the derivatives.

Now let us prove that the solution of the free Maxwell equations (Eq.~(\ref{eq:22})) is normalizable, i.e. integrals of products of any two field components (such as $E_\phi H_\rho$, $E_\rho^2$ etc.) over any transverse section $z=\textrm{const}$ are finite, so energy flow through a transverse section and total electromagnetic energy between two transverse sections are finite. It may be seen from Eq.~(\ref{eq:22}) that all field components are continuous, so it is sufficient to prove that for any $z$ field components are $\textrm{O}(\rho^{-\frac{3}{2}})$ for large $\rho$. Eq.~(\ref{eq:37}) implies that it is sufficient to prove this for components of electric field. We obtain from Eq.~(\ref{eq:65}):
\begin{widetext}
\begin{eqnarray}\label{eq:673}
\bm{E}(\rho,\varphi,z)=w_0^2\int_0^1\exp\left(i\varphi+i\sqrt{1-\lambda^2}\,z-w_0^2\frac{\lambda^2}{2}\right)
\nonumber\\
\times\left\{\frac{i}{\rho}J_1(\lambda\rho)+i\sqrt{1-\lambda^2}\,\lambda J'_1(\lambda\rho),-\lambda J'_1(\lambda\rho)-\frac{1}{\rho}\sqrt{1-\lambda^2}J_1(\lambda\rho),-\frac{1}{\rho}\lambda J'_1(\lambda\rho)-\lambda^2 J''_1(\lambda\rho)+\frac{1}{\rho^2}J_1(\lambda\rho)\right\}\,d\lambda.
\end{eqnarray}
\end{widetext}
If we denote $\exp\left(i\varphi+i\sqrt{1-\lambda^2}\,z-w_0^2\frac{\lambda^2}{2}\right)$ as $q$, then
\begin{eqnarray}\label{eq:674}
|q|\leq 1,
\nonumber\\
|J_0(\lambda\rho)|\leq 1,
\nonumber\\
|J_1(\lambda\rho)|\leq 1.
\end{eqnarray}
Therefore, the term with $\rho^{-2}$ is $\textrm{O}(\rho^{-2})$ for large $\rho$. Let us now assess terms with $\rho^{-1}$, using the first term as an example:
\begin{eqnarray}\label{eq:675}
\int_0^1 q J_1(\lambda\rho)\,d\lambda=\int_0^1 q \frac{-d(J_0(\lambda\rho))}{\rho}=
\nonumber\\
=\left[-\frac{1}{\rho}q J_0(\lambda\rho)\right]_0^1+\frac{1}{\rho}\int_0^1 J_0(\lambda\rho)\,d q=\textrm{O}(\rho^{-1})
\nonumber\\
+\frac{1}{\rho}\int_0^1 q J_0(\lambda\rho)\left(\frac{-i \lambda z}{\sqrt{1-\lambda^2}}-w_0^2\lambda\right)\,d\lambda=\textrm{O}(\rho^{-1})
\end{eqnarray}
(the singularity in the last integral is integrable). The other two terms with $\rho^{-1}$ are assessed in a similar way.

Let us assess another term:
\begin{widetext}
\begin{eqnarray}\label{eq:676}
\int_0^1 q \sqrt{1-\lambda^2}\,\lambda J'_1(\lambda\rho)\,d\lambda=\int_0^1 q \sqrt{1-\lambda^2}\,\lambda \frac{d(J_1(\lambda\rho))}{\rho}=
\nonumber\\
=-\frac{1}{\rho}\int_0^1 J_1(\lambda\rho)q
\left(\left(\frac{-i \lambda z}{\sqrt{1-\lambda^2}}-w_0^2\lambda\right)\sqrt{1-\lambda^2}\,\lambda-\frac{\lambda^2}{\sqrt{1-\lambda^2}}+\sqrt{1-\lambda^2}\right)\,d \lambda.
\end{eqnarray}
\end{widetext}
The last integral may be presented as a sum of two integrals:
\begin{equation}\label{eq:677}
\int_0^1=\int_0^{\frac{1}{2}}+\int_{\frac{1}{2}}^1.
\end{equation}
The first integral may be assessed using another integration by parts, similarly to the previous integrals (this integral does not contain any singularities). The second integral is $\textrm{O}(\rho^{-\frac{1}{2}})$, as for $\lambda\geq\frac{1}{2}$ we have $J_1(\lambda\rho)=\textrm{O}(\rho^{-\frac{1}{2}})$. Thus, the whole term is $\textrm{O}(\rho^{-\frac{3}{2}})$.

The remaining two terms can be assessed in a similar way.
\maketitle

\section{\label{sec:level13}The exact solution for the problem of diffraction on an infinite homogeneous cylinder}

We shall extensively use the exact solution of the problem of diffraction on an infinite homogeneous cylinder by J. Wait Ref.~\cite{Wait1}, so let us outline it here. Let us consider cylindrical waves of TM type:
\begin{eqnarray}\label{eq:73}
\Pi=Z_n(\lambda\rho)\exp(i(\gamma z+n\varphi)),
\nonumber\\
\bm{H}=\left\{\frac{n k_0}{\rho}Z_n(\lambda\rho),i k_0\lambda
Z'_n(\lambda\rho),0\right\}\exp(i(\gamma z+n\varphi)),
\nonumber\\
\bm{E}=\left\{\frac{i \gamma\lambda}{\varepsilon}Z'_n(\lambda\rho),-\frac{\gamma n}{\varepsilon\rho}Z_n(\lambda\rho),\frac{\lambda^2}{\varepsilon}Z_n(\lambda\rho)\right\}
\nonumber\\
\times\exp(i(\gamma z+n\varphi)),\quad
\end{eqnarray}
and TE type:
\begin{eqnarray}\label{eq:74}
\Pi'=Z_n(\lambda\rho)\exp(i(\gamma z+n\varphi)),
\nonumber\\
\bm{H'}=\left\{\frac{i \gamma\lambda}{\mu}Z'_n(\lambda\rho),-\frac{\gamma n}{\mu\rho}Z_n(\lambda\rho),\frac{\lambda^2}{\mu}Z_n(\lambda\rho)\right\}
\nonumber\\
\times\exp(i(\gamma z+n\varphi)),
\nonumber\\
\bm{E'}=\left\{-\frac{n k_0}{\rho}Z_n(\lambda\rho),-i k_0\lambda
Z'_n(\lambda\rho),0\right\}\exp(i(\gamma z+n\varphi)),\quad
\end{eqnarray}
where $Z_n$ is a cylindrical function ($J_n$ or $H_n^{(1)}$), and
\begin{equation}\label{eq:75}
\lambda^2=k^2-\gamma^2.
\end{equation}
The cylinder is determined by the equation $\rho\leq a$, so $a$ is the radius of the cylinder, the permittivity and the magnetic permeability are $\varepsilon_1$ and $\mu_1$ outside the cylinder and $\varepsilon_2$ and $\mu_2$ inside the cylinder. Let the incident field be a linear combination of cylindrical TM and TE waves with the following $z$-components of the electric and magnetic Hertz vectors, correspondingly:
\begin{eqnarray}\label{eq:76}
u_0=\alpha J_n(\lambda_1\rho)F_n,
\nonumber\\
v_0=\beta J_n(\lambda_1\rho)F_n,
\end{eqnarray}
where $F_n=\exp(i(\gamma z+n\varphi))$. Then the reflected field
is also a linear combination of cylindrical TM and TE waves:
\begin{eqnarray}\label{eq:77}
u_1=a_{1,n} H_n^{(1)}(\lambda_1\rho)F_n,
\nonumber\\
v_1=b_{1,n} H_n^{(1)}(\lambda_1\rho)F_n,
\end{eqnarray}
and the same is true for the refracted field:
\begin{eqnarray}\label{eq:78}
u_2=a_{2,n} J_n(\lambda_2\rho)F_n,
\nonumber\\
v_2=b_{2,n} J_n(\lambda_2\rho)F_n,
\end{eqnarray}
($u_1$ and $u_2$ are the $z$-components of the electric Hertz
vector, $v_1$ and $v_2$ are the $z$-components of the magnetic Hertz
vector). The coefficients $a_{1,n}$, $a_{2,n}$, $b_{1,n}$,
$b_{2,n}$ are evaluated from the continuity conditions for the
tangential components of the fields on the surface of the
cylinder. We shall omit subscripts $n$ and superscripts $(1)$ for simplicity. For
$\rho=a$ we have (superscripts $inc$, $r\!fl$, and $r\!fr$ relate
to the incident, reflected, and refracted fields; $\lambda_1
a=p_1$, $\lambda_2 a=p_2$):
\begin{eqnarray}\label{eq:79}
H_{\varphi}^{inc}+H_{\varphi}^{r\!fl}=H_{\varphi}^{r\!fr},
\nonumber\\
\alpha i k_0\lambda_1 J'(p_1)-\beta\frac{\gamma n}{a
\mu_1}J(p_1)+a_1 i k_0\lambda_1 H'(p_1)-
\nonumber\\
-b_1\frac{\gamma n}{a \mu_1}H(p_1)=i k_0 \lambda_2 a_2
J'(p_2)-\frac{\gamma n}{a \mu_2}b_2 J(p_2),
\end{eqnarray}
\begin{eqnarray}\label{eq:80}
E_{\varphi}^{inc}+E_{\varphi}^{r\!fl}=E_{\varphi}^{r\!fr},
\nonumber\\
-\frac{\gamma n}{\varepsilon_1 a}\alpha J(p_1)-i k_0
\lambda_1\beta J'(p_1)-\frac{\gamma n}{\varepsilon_1 a}a_1 H(p_1)-
\nonumber\\
-i k_0 \lambda_1 b_1 H'(p_1)=-\frac{\gamma n}{\varepsilon_2 a}a_2
J(p_2)-i k_0\lambda_2b_2 J'(p_2),
\end{eqnarray}
\begin{eqnarray}\label{eq:81}
H_{z}^{inc}+H_{z}^{r\!fl}=H_{z}^{r\!fr},
\nonumber\\
\frac{\lambda_1^2}{\mu_1}\beta J(p_1)+\frac{\lambda_1^2}{\mu_1}b_1
H(p_1)=\frac{\lambda_2^2}{\mu_2}b_2 J(p_2),
\end{eqnarray}
\begin{eqnarray}\label{eq:82}
E_{z}^{inc}+E_{z}^{r\!fl}=E_{z}^{r\!fr},
\nonumber\\
\frac{\lambda_1^2}{\varepsilon_1}\alpha
J(p_1)+\frac{\lambda_1^2}{\varepsilon_1}a_1
H(p_1)=\frac{\lambda_2^2}{\varepsilon_2}a_2 J(p_2).
\end{eqnarray}
We obtain:
\begin{eqnarray}\label{eq:83}
b_1=\frac{\mu_1}{\lambda_1^2
H(p_1)}\left(\frac{\lambda_2^2}{\mu_2} b_2
J(p_2)-\frac{\lambda_1^2}{\mu_1}\beta J(p_1)\right)=
\nonumber\\
=\frac{\mu_1}{\mu_2}\frac{\lambda_2^2}{\lambda_1^2}\frac{J(p_2)}{H(p_1)}b_2
-\frac{J(p_1)}{H(p_1)}\beta,
\end{eqnarray}
\begin{eqnarray}\label{eq:84}
a_1=\frac{\varepsilon_1}{\lambda_1^2
H(p_1)}\left(\frac{\lambda_2^2}{\varepsilon_2} a_2
J(p_2)-\frac{\lambda_1^2}{\varepsilon_1}\alpha J(p_1)\right)=
\nonumber\\
=\frac{\varepsilon_1}{\varepsilon_2}\frac{\lambda_2^2}{\lambda_1^2}\frac{J(p_2)}{H(p_1)}a_2
-\frac{J(p_1)}{H(p_1)}\alpha.
\end{eqnarray}
Let us exclude $a_1$ and $b_1$ from
Eqs.~(\ref{eq:79}-\ref{eq:80}):
\begin{eqnarray}\label{eq:85}
\alpha i k_0\lambda_1 J'(p_1)-\beta\frac{\gamma n}{a \mu_1}J(p_1)+
\nonumber\\
+\left(\frac{\varepsilon_1}{\varepsilon_2}\frac{\lambda_2^2}{\lambda_1^2}\frac{J(p_2)}{H(p_1)}a_2
-\frac{J(p_1)}{H(p_1)}\alpha\right) i k_0\lambda_1 H'(p_1)-
\nonumber\\
-\left(\frac{\mu_1}{\mu_2}\frac{\lambda_2^2}{\lambda_1^2}\frac{J(p_2)}{H(p_1)}b_2
-\frac{J(p_1)}{H(p_1)}\beta\right)\frac{\gamma n}{a \mu_1}H(p_1)=
\nonumber\\
=i k_0 \lambda_2 a_2 J'(p_2)-\frac{\gamma n}{a \mu_2}b_2 J(p_2),
\end{eqnarray}
\begin{eqnarray}\label{eq:86}
-\frac{\gamma n}{\varepsilon_1 a}\alpha J(p_1)-i k_0 \lambda_1\beta J'(p_1)-
\nonumber\\
-\frac{\gamma n}{\varepsilon_1
a}\left(\frac{\varepsilon_1}{\varepsilon_2}\frac{\lambda_2^2}{\lambda_1^2}\frac{J(p_2)}{H(p_1)}a_2
-\frac{J(p_1)}{H(p_1)}\alpha\right) H(p_1)-
\nonumber\\
-i k_0 \lambda_1
\left(\frac{\mu_1}{\mu_2}\frac{\lambda_2^2}{\lambda_1^2}\frac{J(p_2)}{H(p_1)}b_2
-\frac{J(p_1)}{H(p_1)}\beta\right) H'(p_1)=
\nonumber\\
=-\frac{\gamma n}{\varepsilon_2 a}a_2 J(p_2)-i k_0\lambda_2b_2
J'(p_2).
\end{eqnarray}
Thus, we have a system of two linear equations for $a_2$ and
$b_2$:
\begin{equation}\label{eq:87}
c_{11}a_2+c_{12}b_2=c_1,
\end{equation}
\begin{equation}\label{eq:88}
c_{21}a_2+c_{22}b_2=c_2,
\end{equation}
where
\begin{eqnarray}\label{eq:89}
c_{11}=\left(\frac{1}{p_{1}}\frac{H'(p_{1})}{H(p_{1})}-
\frac{1}{p_{2}}\frac{\varepsilon_{2}}{\varepsilon_{1}}\frac{J'(p_{2})}{J(p_{2})}\right)
\frac{\varepsilon_{1}p_{2}}{\varepsilon_{2}}ik_{0}\frac{p_{2}}{a}J(p_{2}),
\nonumber\\
c_{12}=\left(\frac{1}{p^{2}_{1}}-\frac{1}{p^{2}_{2}}\right)(-p^{2}_{2})\frac{\gamma
n}{a\mu_{2}}J(p_{2}),
\nonumber\\
c_{21}=\left(\frac{1}{p^{2}_{1}}-\frac{1}{p^{2}_{2}}\right)(-p^{2}_{2})\frac{\gamma
n}{a\varepsilon_{2}}J(p_{2}),
\nonumber\\
c_{22}=\left(\frac{1}{p_{1}}\frac{H'(p_{1})}{H(p_{1})}-
\frac{1}{p_{2}}\frac{\mu_{2}}{\mu_{1}}\frac{J'(p_{2})}{J(p_{2})}\right)
\nonumber\\
\times\frac{(-\mu_{1}p_{2})}{\mu_{2}}ik_{0}\frac{p_{2}}{a}J(p_{2}),
\nonumber\\
c_1=\alpha i
k_{0}\lambda_{1}\left(\frac{H'(p_{1})}{H(p_{1})}-\frac{J'(p_{1})}{J(p_{1})}\right)J(p_{1})=
\nonumber\\
=\frac{\alpha i k_{0}\lambda_{1}}{H(p_{1})J(p_{1})}\frac{2i}{\pi
p_{1}}J(p_{1}),
\nonumber\\
c_2=i k_0\lambda_1\beta
J(p_1)\left(\frac{J'(p_{1})}{J(p_{1})}-\frac{H'(p_{1})}{H(p_{1})}\right)=
\nonumber\\
=\frac{i k_0\lambda_1\beta J(p_1)}{J(p_1)H(p_1)}\left(\frac{-2
i}{\pi p_1}\right).\hspace{6pt}
\end{eqnarray}
We have used the well-known formula for the Wronskian of cylindrical
functions:
\begin{equation}\label{eq:90}
W[J_n(z),H_n(z)]=\frac{2 i}{\pi z}.
\end{equation}
Let us solve the system using the Cramer's Rule.
\begin{eqnarray}\label{eq:91}
D'=\left|\begin{array}{cc}
  c_{11} & c_{12} \\
  c_{21} & c_{22}
\end{array}\right|=\frac{\varepsilon_1}{\varepsilon_2}
\frac{i k_0 p^2_2 J(p_2)}{a}\frac{\mu_1}{\mu_2}\frac{i k_0 p^2_2
J(p_2)}{a}D,
\end{eqnarray}
where
\begin{widetext}
\begin{eqnarray}\label{eq:92}
D=\left|\begin{array}{cc}
  \frac{1}{p_{1}}\frac{H'(p_{1})}{H(p_{1})}-
\frac{1}{p_{2}}\frac{\varepsilon_{2}}{\varepsilon_{1}}\frac{J'(p_{2})}{J(p_{2})} &
 -\frac{\gamma n}{i k_0 \mu_1} \left(\frac{1}{p^2_1}-\frac{1}{p^2_2}\right)\\
  -\frac{\gamma n}{i k_0 \varepsilon_1} \left(\frac{1}{p^2_1}-\frac{1}{p^2_2}\right) &
   -\left(\frac{1}{p_{1}}\frac{H'(p_{1})}{H(p_{1})}-
\frac{1}{p_{2}}\frac{\mu_{2}}{\mu_{1}}\frac{J'(p_{2})}{J(p_{2})}\right)
\end{array}  \right|=\frac{\gamma^2
n^2}{k^2_1}\left(\frac{1}{p^2_1}-
\frac{1}{p^2_2}\right)^2-
\nonumber\\
-\left(\frac{1}{p_{1}}\frac{H'(p_{1})}{H(p_{1})}-
\frac{1}{p_{2}}\frac{\varepsilon_{2}}{\varepsilon_{1}}\frac{J'(p_{2})}{J(p_{2})}\right)
\left(\frac{1}{p_{1}}\frac{H'(p_{1})}{H(p_{1})}-
\frac{1}{p_{2}}\frac{\mu_{2}}{\mu_{1}}\frac{J'(p_{2})}{J(p_{2})}\right).
\end{eqnarray}
\end{widetext}
Then we have: \begin{widetext}
\begin{eqnarray}\label{eq:93}
D_1=\left|\begin{array}{cc}
  c_{1} & c_{12} \\
  c_{2} & c_{22}
\end{array}\right|=\left|\begin{array}{cc}
  -\frac{2\alpha}{a\pi}\frac{k_0}{H(p_1)} & -\frac{\gamma n}{i k_0 \mu_1}\left(\frac{1}{p_1^2}-\frac{1}{p_2^2}\right) \\
  \frac{2\beta}{a\pi}\frac{k_0}{H(p_1)} & -\left(\frac{1}{p_{1}}\frac{H'(p_{1})}{H(p_{1})}-
\frac{1}{p_{2}}\frac{\mu_{2}}{\mu_{1}}\frac{J'(p_{2})}{J(p_{2})}\right)
\end{array}\right|\frac{\mu_1}{\mu_2}i k_0\frac{p^2_2 J(p_2)}{a}=
\nonumber\\
=\left|\begin{array}{cc}
  -\alpha & -\frac{\gamma n}{i k_0 \mu_1}\left(\frac{1}{p_1^2}-\frac{1}{p_2^2}\right) \\
  \beta & -\left(\frac{1}{p_{1}}\frac{H'(p_{1})}{H(p_{1})}-
\frac{1}{p_{2}}\frac{\mu_{2}}{\mu_{1}}\frac{J'(p_{2})}{J(p_{2})}\right)
\end{array}\right|\frac{2 k_0}{a \pi H(p_1)}\frac{\mu_1}{\mu_2}i k_0\frac{p^2_2
J(p_2)}{a},
\end{eqnarray}
hence,
\begin{eqnarray}\label{eq:94}
a_2=\frac{D_1}{D'}=\frac{-2 i}{\pi H(p_1)
D}\frac{\varepsilon_2}{\varepsilon_1}\frac{1}{p_2^2
J(p_2)}\left(\alpha\left(\frac{1}{p_1}\frac{H'(p_1)}{H(p_1)}-
\frac{1}{p_2}\frac{\mu_2}{\mu_1}\frac{J'(p_2)}{J(p_2)}\right)-\beta\frac{i
\gamma
n}{k_0\mu_1}\left(\frac{1}{p_1^2}-\frac{1}{p_2^2}\right)\right)=
\nonumber\\
=\alpha
\frac{\varepsilon_2}{\varepsilon_1}\frac{1}{J(p_2)}\frac{-2 i}{\pi
p_2^2 H(p_1)D}\left(\frac{1}{p_1}\frac{H'(p_1)}{H(p_1)}-
\frac{1}{p_2}\frac{\mu_2}{\mu_1}\frac{J'(p_2)}{J(p_2)}\right)+
\beta\frac{\varepsilon_2}{\varepsilon_1}\frac{1}{J(p_2)}\frac{-2
}{\pi p_2^2 H(p_1)D}\frac{\gamma n}{k_0
\mu_1}\left(\frac{1}{p_1^2}-\frac{1}{p_2^2}\right).
\end{eqnarray}
Similarly,
\begin{eqnarray}\label{eq:95}
D_2=\left|\begin{array}{cc}
  c_{11} & c_{1} \\
  c_{21} & c_{2}
\end{array}\right|=\left|\begin{array}{cc}
  \frac{1}{p_1}\frac{H'(p_1)}{H(p_1)}-
\frac{1}{p_2}\frac{\varepsilon_2}{\varepsilon_1}\frac{J'(p_2)}{J(p_2)} & -\alpha \\
  -\frac{\gamma n}{i k_0 \varepsilon_1}\left(\frac{1}{p_1^2}-\frac{1}{p_2^2}\right) & \beta
\end{array}\right|
\frac{\varepsilon_1}{\varepsilon_2}\frac{i k_0 p_2^2
J(p_2)}{a}\frac{2 k_0}{a\pi H(p_1)},
\end{eqnarray}
\begin{eqnarray}\label{eq:96}
b_2=\frac{D_2}{D'}=\frac{-2 i}{\pi H(p_1)
D}\frac{\mu_2}{\mu_1}\frac{1}{p_2^2 J(p_2)}\left(\alpha
\frac{i\gamma n}{k_0\varepsilon_1}\left(\frac{1}{p_1^2}-
\frac{1}{p_2^2}\right)+
\beta\left(\frac{1}{p_1}\frac{H'(p_1)}{H(p_1)}-
\frac{1}{p_2}\frac{\varepsilon_2}{\varepsilon_1}\frac{J'(p_2)}{J(p_2)}\right)\right)=
\nonumber\\
 =\alpha\frac{\mu_2}{\mu_1} \frac{1}{J(p_2)}
\frac{2}{\pi p_2^2 H(p_1)D}\frac{\gamma
n}{k_0\varepsilon_1}\left(\frac{1}{p_1^2}-\frac{1}{p_2^2}\right)+\beta
\frac{\mu_2}{\mu_1}\frac{1}{J(p_2)}\frac{-2 i}{\pi p_2^2
H(p_1)D}\left(\frac{1}{p_1}\frac{H'(p_1)}{H(p_1)}-
\frac{1}{p_2}\frac{\varepsilon_2}{\varepsilon_1}\frac{J'(p_2)}{J(p_2)}\right).
\end{eqnarray}
\end{widetext}
\section{\label{sec:level14}The conditions of efficient absorption of a broad
Gaussian electromagnetic beam in a thin conducting cylinder}
Diffraction problems are often highly sensitive to variations of
their parameters. Therefore, confirmation of the results of any
calculations by rigorous solution of Maxwell equations is highly
desirable, and this procedure is discussed in the next section.
However, one also needs simpler approximate procedures to make
estimates and to determine the most promising parameter domains.
Therefore, in this section the author describes a simple (and
sometimes inaccurate) qualitative estimate of the part of the
power of a Gaussian beam absorbed in a conducting cylinder and a
semiquantitative one-wave method. The latter is used to deduce an
asymptotic formula for efficiency of the absorption and to
determine the parameter domains where efficient absorption takes
place. The results of the semiquantitative methods were confirmed
by rigorous solution of the Maxwell equation. This section mostly
deals with the longitudinal geometry, as for this geometry
conditions of efficient absorption of a broad beam in a thin
cylinder were not known, so the author investigates this case in
much detail. For the transverse geometry some resonant cases of
efficient absorption were known (see, eg.,Ref.~\cite{Zharov1}), so
the author describes just one non-resonant parameter domain where
efficient absorption takes place in a cylinder with a radius that
is much less than the wavelength. This result is important as it
complements the results for the longitudinal geometry. To describe
the qualitative estimate for the longitudinal geometry, let us
consider a Gaussian beam of electromagnetic waves with frequency
$\omega$ in free space. Let the length of the conducting cylinder
be greater than the beam waist length $l_1$ defined by twofold
field intensity attenuation, its radius $a$ be (much) less than
that of the beam waist $r_1$ defined by $e$-fold field intensity
attenuation, and its conductivity $\sigma$ be such that the
skin-depth be roughly equal to the cylinder diameter:
\begin{equation}\label{eq:97}
4a^2=\frac{c^2}{2\pi\omega\sigma}.
\end{equation}
Let us place the cylinder in the beam in such a way that the axes
of the cylinder and the beam coincide, and the cylinder overlap
the whole beam waist length. Let us assume that the electric field
within the cylinder is equal in absolute value to that in the
incident wave, i.e. that in the absence of the cylinder (the
field penetrates the conducting medium to skin-depth). In general, this
assumption is wrong and cannot be defended. However, in some
important cases it does hold true and allows to obtain a simple
estimate and understand the scaling of the problem. Let us also
assume that within the beam waist the electric field in the
incident wave is equal in absolute value to that in the center of
the beam waist $\textbf{E}$ (this assumption cannot lead to
additional overvaluation by more than $e$ times). The beam power
equals
\begin{equation}\label{eq:98}
W=S_1|\textbf{P}|=\pi
r_1^2\cdot\frac{1}{2}\cdot\frac{c}{4\pi}\textbf{E}^2=\frac{1}{8}c r_1^2
\textbf{E}^2.
\end{equation}
Here $S_1$ is the area of the transverse section of the beam
waist, and $|\textbf{P}|$ is the averaged Poynting vector in the
beam waist. The power absorbed in the cylinder equals
\begin{equation}\label{eq:99}
W^a=V\cdot\frac{1}{2}\sigma \textbf{E}^2=\pi
a^2\cdot 2\frac{\omega}{c}
r_1^2\cdot\frac{1}{2}\cdot\frac{c^2 \textbf{E}^2}{2
\pi\omega\cdot 4 a^2}=\frac{1}{8}c r_1^2 \textbf{E}^2.
\end{equation}
Here $V=\pi a^2 l_1$ is the volume where absorption takes place,
\begin{equation}\label{eq:99a}
l_1=2\frac{\omega}{c}
r_1^2
\end{equation}
is the beam waist length for the Gaussian beam
(Ref.~\cite{Shimoda1}). Thus, the absorption efficiency
$\frac{W_a}{W}$ equals unity to the order of magnitude. One can
see that if the beam waist width increases twofold, the electric
field in the cylinder falls fourfold (if beam power does not change), but the beam waist length
increases fourfold, so the absorption efficiency does not change.
Alternatively, if we use a cylinder with the radius that is half
as large, the optimum conductivity increases fourfold, so in this
case the absorption efficiency does not change, either. However,
this estimate calculation is not always correct, so let us
describe a semiquantitative method of calculation of the
absorption efficiency. This is actually a one-wave approximation.
The idea is as follows. Any incident beam can be expanded in
cylindrical waves (see Eqs.~(\ref{eq:73}) and (\ref{eq:74})). For
each cylindrical wave there is a relatively simple exact solution
of the problem of diffraction on an infinite cylinder. Let us
choose a typical cylindrical wave from the expansion. Its field on
the axis will have a structure similar to that of the incident
beam. We may then assume that for the incident beam within the
beam waist length the refracted wave coincides with that for the
chosen cylindrical wave (the latter should be normalized in such a
way that the magnitudes of the fields in the incident beam and in
this wave are approximately equal).
So let us assume that the actual incident, reflected, and refracted fields may be approximated within the beam waist by Eqs.~(\ref{eq:76}),(\ref{eq:77}), and (\ref{eq:78}), correspondingly. We further assume that subscript $n=1$ (otherwise there is no efficient absorption in the longitudinal geometry, at least if $\lambda_1\ll 1$, which is true for cylindrical waves that have a field structure similar to that in the waist of Gaussian beams), and this subscript will be omitted. Let us choose
\begin{equation}\label{eq:99b}
\lambda_1\approx\frac{2}{r_1},
\end{equation}
so that the width of the Gaussian beam waist is approximately equal to the width of the main maximum of function $J_0(\lambda_1\rho)$.
We are interested in analytical conditions of efficient absorption of broad beams in thin cylinders, so we make the following assumptions:
\begin{equation}\label{eq:100}
p_{1}\ll 1,
\end{equation}
\begin{equation}\label{eq:101}
\frac{1}{p^{2}_1}\gg \left|\frac{1}{p_2}\frac{J'(p_2)}{J(p_2)}\right|,
\end{equation}
\begin{equation}\label{eq:102}
\frac{1}{p^{2}_1}\gg \left|\frac{\varepsilon_2}{\varepsilon_1}\frac{1}{p_2}\frac{J'(p_2)}{J(p_2)}\right|,
\end{equation}
\begin{equation}\label{eq:103}
|p_2|\gg p_1,
\end{equation}
\begin{equation}\label{eq:104}
\lambda_1\ll 1.
\end{equation}
Eq.~(\ref{eq:104}) implies
\begin{equation}\label{eq:105}
\gamma\approx 1.
\end{equation}
From here on we assume that
\begin{equation}\label{eq:105a}
\mu_1=\mu_2=\varepsilon_1=1
\end{equation}
(the ambient space has the properties of vacuum, and the material of the cylinder is non-magnetic) and omit the subscript in $\varepsilon_2$.
Then, taking into account that
\begin{equation}\label{eq:106}
H^{(1)}_1(x)\approx-\frac{2i}{\pi x},
\end{equation}
\begin{equation}\label{eq:107}
\frac{H^{(1)'}_1(x)}{H^{(1)}_1(x)}\approx-\frac{1}{x}
\end{equation}
for $x\ll 1$, we obtain:
\begin{eqnarray}\label{eq:108}
a_2\approx\alpha\varepsilon\frac{1}{J(p_2)}\frac{-2i}{\pi p^2_2\left(-\frac{2i}{\pi p_1}\right)D}\left(-\frac{1}{p^2_1}\right)+
\nonumber\\
+\beta\varepsilon\frac{1}{J(p_2)}\frac{-2}{\pi p^2_2\left(-\frac{2i}{\pi p_1}\right)D}\frac{1}{p^2_1}=\alpha\frac{i}{p^2_2 J(p_2) p_1D}i\varepsilon+
\nonumber\\
+\beta\frac{i}{p^2_2 J(p_2) p_1D}(-\varepsilon)=\frac{i}{p^2_2 J(p_2) p_1D}i\varepsilon(\alpha+i\beta),
\nonumber\\
b_2\approx\alpha\frac{1}{J(p_2)}\frac{2}{\pi p^2_2\left(-\frac{2i}{\pi p_1}\right)D}\frac{1}{p^2_1}+
\nonumber\\
+\beta\frac{1}{J(p_2)}\frac{-2i}{\pi p^2_2\left(-\frac{2i}{\pi p_1}\right)D}\left(-\frac{1}{p^2_1}\right)=\alpha\frac{i}{p^2_2 J(p_2) p_1 D}+
\nonumber\\
+\beta\frac{i}{p^2_2 J(p_2) p_1 D}i=\frac{i}{p^2_2 J(p_2) p_1D}(\alpha+i\beta).\hspace{12pt}
\end{eqnarray}
Let us estimate power in the incident beam. If complex amplitudes of electric and magnetic fields $\bm{E}_{am}$ and $\bm{H}_{am}$ in the center of the beam waist ($\rho=0, z=0$) are the same as those for the linear combination of cylindrical waves of Eq.~(\ref{eq:76}), then we obtain using Eqs.~(\ref{eq:73}),(\ref{eq:74}):
\begin{eqnarray}\label{eq:109}
\bm{E}_{am}\approx(\alpha+i\beta)\frac{\lambda_1}{2}\{i,-1,0\},
\nonumber\\
\bm{H}_{am}\approx(\alpha+i\beta)\frac{\lambda_1}{2}\{1,i,0\}
\end{eqnarray}
(we take into account that
\begin{eqnarray}\label{eq:110}
\lim_{\rho \to 0}\frac{J_1(\lambda_1\rho)}{\rho}=\frac{\lambda_1}{2},
\nonumber\\
J'_1(0)=\frac{1}{2},
\end{eqnarray}
use Eqs.~(\ref{eq:104}),(\ref{eq:105}), and omit factor $\exp(i n\varphi)$).
The $z$-component of the Poynting vector in the center of the incident beam averaged over a period equals
\begin{eqnarray}\label{eq:111}
\frac{1}{2}\frac{c}{4\pi} \textrm{Re}[\bm{E}_{am}\times\bm{H}^{*}_{am}]_{z}=
\nonumber\\
=\frac{1}{2}\frac{c}{4\pi} \textrm{Re}(E_{am\rho}H^{*}_{am\varphi}-E_{am\varphi}H^{*}_{am\rho})=
\nonumber\\
=\frac{1}{2}\frac{c}{4\pi} \textrm{Re}((\alpha+i\beta)(a^*-i\beta^*)\frac{\lambda^{2}_1}{4}\cdot 2=
\theta\frac{c\lambda^2_1}{16\pi},
\end{eqnarray}
where $\theta=\alpha\alpha^*+\beta\beta^*+i\alpha^*\beta-i\alpha\beta^*$.
We may assume that the effective area of the transverse section of the Gaussian beam waist equals
\begin{equation}\label{eq:112}
\int dx dy \exp(-\frac{\rho^2}{r_1^2})=\pi r_1^2,
\end{equation}
where $r_1$ is the radius of the beam waist, then power in the incident beam equals
\begin{equation}\label{eq:113}
W=\theta\frac{c\lambda^2_1}{16\pi}\pi r_1^2=\theta\frac{c\lambda_1^2 r_1^2}{16}=\theta\frac{c p_1^2 r_1^2}{16 a^2}.
\end{equation}
Let us estimate the energy flow through the lateral surface of the cylinder. The $\rho$-component of the Poynting vector at $z=0$, $\rho=a$ averaged over a period equals
\begin{equation}\label{eq:114}
\frac{1}{2}\frac{c}{4\pi} \textrm{Re}[\bm{E}^{r\!fr}_{a}(a,\varphi,0)\times\bm{H}^{r\!fr*}_{a}(a,\varphi,0)]_{\rho}.
\end{equation}
From Eqs.~(\ref{eq:73}),(\ref{eq:74}), and (\ref{eq:78}) we find for  complex amplitudes of the fields in the refracted beam:
\begin{eqnarray}\label{eq:115}
\bm{H}^{r\!fr}_{a}(a,\varphi,0)=a_2 \bm{H}_0+b_2 \bm{H'}_0,
\nonumber\\
\bm{E}^{r\!fr}_{a}(a,\varphi,0)=a_2 \bm{E}_0+b_2 \bm{E'}_0,
\end{eqnarray}
where
\begin{eqnarray}\label{eq:116}
\bm{H}_0=\frac{1}{a}\left\{J(p_2),ip_2 J'(p_2),0\right\},
\nonumber\\
\bm{E}_0=\frac{1}{a}\left\{\frac{i\gamma p_2}{\varepsilon}J'(p_2),-\frac{\gamma}{\epsilon} J(p_2),\frac{p_2^2}{\varepsilon a}J(p_2)\right\},
\nonumber\\
\bm{H'}_0=\frac{1}{a}\left\{i\gamma p_2 J'(p_2),-\gamma J(p_2),\frac{p_2^2}{a}J(p_2)\right\},
\nonumber\\
\bm{E'}_0=\frac{1}{a}\left\{-J(p_2),-ip_2 J'(p_2),0\right\}
\end{eqnarray}
(we omit factor $\exp(i n\varphi)$). Therefore, in view of Eq.~(\ref{eq:105}),
\begin{eqnarray}\label{eq:117}
E^{r\!fr}_{a\varphi}=E^{r\!fr}_{a\varphi}(a,\varphi,0)=
\nonumber\\
=(\alpha+i\beta)\frac{i}{a p_2^2 J p_1 D}\left(i\varepsilon\left(-\frac{1}{\varepsilon}\right)J+(-ip_2 J')\right)=
\nonumber\\
=(\alpha+i\beta)\frac{i}{a p_2^2 J p_1 D}(-i)(J+p_2 J'),
\nonumber\\
E^{r\!fr}_{a z}=E^{r\!fr}_{a z}(a,\varphi,0)=
\nonumber\\
=(\alpha+i\beta)\frac{i}{a p_2^2 J p_1 D}i\varepsilon\frac{p_2^2}{\varepsilon a}J=
\nonumber\\
=(\alpha+i\beta)\frac{i}{a p_2^2 J p_1 D}\frac{i p_2^2}{a}J,
\nonumber\\
H^{r\!fr}_{a\varphi}=H^{r\!fr}_{a\varphi}(a,\varphi,0)=
\nonumber\\
=(\alpha+i\beta)\frac{i}{a p_2^2 J p_1 D}(i\varepsilon\ i p_2 J'+(-J))=
\nonumber\\
=(\alpha+i\beta)\frac{i}{a p_2^2 J p_1 D}(-1)(J+\varepsilon p_2 J'),
\nonumber\\
H^{r\!fr}_{a z}=H^{r\!fr}_{a z}(a,\varphi,0)=
\nonumber\\
=(\alpha+i\beta)\frac{i}{a p_2^2 J p_1 D}\frac{p_2^2}{a}J,\hspace{6pt}
\end{eqnarray}
where $J=J_1(p_2)$, $J'=J'_1(p_2)$.
\begin{eqnarray}\label{eq:118}
\textrm{Re}[\bm{E}^{r\!fr}_{a}(a,\varphi,0)\times\bm{H}^{r\!fr*}_{a}(a,\varphi,0)]_{\rho}=
\nonumber\\
=\textrm{Re}(E^{r\!fr*}_{a\varphi}H^{r\!fr}_{a z}-E^{r\!fr}_{a z}H^{r\!fr*}_{a\varphi})=
\nonumber\\
=(\alpha+i\beta)(a^*-i\beta^*)\frac{1}{a^2 |p_2|^4 |J|^2 p_1^2 |D|^2}\times
\nonumber\\
\times \textrm{Re}\left(i(J^*+p_2^* J'^*)\frac{p_2^2}{a}J-\frac{i p_2^2}{a}J(-1)(J^* +\varepsilon^* p_2^* J'^*)\right)=
\nonumber\\
=\frac{\theta}{a^3 |p_2|^4 p_1^2 |D|^2}\textrm{Im}\left(p_2^2\left(-2-p_2^*\frac{J'^*}{J^*}-\varepsilon^* p_2^*\frac{J'^*}{J^*}\right)\right)=
\nonumber\\
=\frac{\theta}{a^3 |p_2|^4 p_1^2 |D|^2}\textrm{Im}\left(p_2^{*2}\left(2+(\varepsilon+1)p_2\frac{J'}{J}\right)\right)=
\nonumber\\
=\frac{2\theta}{a^3 p_1^2 |D|^2}\textrm{Im}\left(\frac{1}{p_2^2}+\frac{\varepsilon+1}{2}\frac{J'}{p_2 J}\right).\hspace{20pt}
\end{eqnarray}
To estimate the energy flow through the surface of the cylinder within the beam waist, we should multiply the Poynting vector of Eq.~(\ref{eq:113}) by the area of the lateral surface of the cylinder:
\begin{equation}\label{eq:119}
2\pi a l_1
\end{equation}
We neglect the energy flow through the sections of the cylinder at $z=\pm\frac{l_1}{2}$, as in view of Eqs.~(\ref{eq:99a}),(\ref{eq:99b}) it is much less than the power in the incident beam (Eq.~(\ref{eq:113})).
The power absorbed in the cylinder $W^a$ equals the energy flow through the surface of the cylinder with an opposite sign (as positive $\rho$-component of the Poynting vector corresponds to outward energy flow):
\begin{eqnarray}\label{eq:120}
W^a=-2\pi a l_1\frac{1}{2}\frac{c}{4\pi}\frac{2\theta}{a^3 p_1^2 |D|^2}\textrm{Im}\left(\frac{1}{p_2^2}+\frac{\varepsilon+1}{2}\frac{J'}{p_2 J}\right)=
\nonumber\\
=-\frac{\pi k_0}{2\pi}(2 r_1)^2\frac{c}{4}\frac{2\theta}{a^2 p_1^2 |D|^2}\textrm{Im}\left(\frac{1}{p_2^2}+\frac{\varepsilon+1}{2}\frac{J'}{p_2 J}\right)=
\nonumber\\
=-\frac{\theta c r_1^2}{a^2 p_1^2 |D|^2}\textrm{Im}\left(\frac{1}{p_2^2}+\frac{\varepsilon+1}{2}\frac{J'}{p_2 J}\right).\hspace{10pt}
\end{eqnarray}
We used Eqs.~(\ref{eq:99a}),(\ref{eq:118}). From Eqs.~(\ref{eq:113}),(\ref{eq:120}) we obtain an expression for the part of the power of the incident beam that is absorbed in the cylinder:
\begin{eqnarray}\label{eq:121}
\frac{W^a}{W}=-\frac{16}{p_1^4 |D|^2}\textrm{Im}\left(\frac{1}{p_2^2}+\frac{\varepsilon+1}{2}\frac{J'}{p_2 J}\right).
\end{eqnarray}
Let us  estimate determinant $D$ (Eq.~(\ref{eq:92})). We use the assumptions of Eqs.~(\ref{eq:100}-\ref{eq:105a}), but sometimes we need more accurate estimates. Let us introduce the following designations:
\begin{eqnarray}\label{eq:122}
f_1(p_1)=\frac{1}{p_1}\frac{H'(p_1)}{H(p_1)},
\nonumber\\
f_2=f_2(p_2)=\frac{1}{p_2}\frac{J'(p_2)}{J(p_2)}.
\end{eqnarray}
Then
\begin{eqnarray}\label{eq:123}
D=\left(1-\frac{p_1^2}{a^2}\right)\left(\frac{1}{p_1^2}-\frac{1}{p_2^2}\right)^2-
\nonumber\\
-(f_1(p_1)-f_2(p_2))(f_1(p_1)-\varepsilon f_2(p_2)).
\end{eqnarray}
Let us introduce an additional assumption:
\begin{equation}\label{eq:123a}
|\ln(p_1)|\gg 1.
\end{equation}
This assumption may seem unrealistic, as it means that the radius of the cylinder is several orders of magnitude less than the radius of the beam waist. However, this assumption allows to derive asymptotic formulae that give very useful guidelines for more realistic cases. It seems worthwhile to analyze the case of extremely thin cylinders, as they are more difficult to heat.
Thus, let us use the following expansions for the Hankel functions with small arguments:
\begin{eqnarray}\label{eq:124}
H_1^{(1)}(x)\approx\frac{2 i}{\pi}\left(\frac{x}{2}\ln x-\frac{1}{x}\right),
\nonumber\\
H_1^{(1)'}(x)\approx\frac{2 i}{\pi}\left(\frac{1}{2}\ln x+\frac{1}{x^2}\right),
\end{eqnarray}
so
\begin{eqnarray}\label{eq:125}
f_1(p_1)\approx\frac{1}{p_1}\frac{\frac{1}{2}\ln p_1+\frac{1}{p_1^2}}{\frac{p_1}{2}\ln p_1-\frac{1}{p_1}}=-\frac{1}{p_1^2}\frac{1+\frac{1}{2}p_1^2\ln p_1}{1-\frac{1}{2}p_1^2\ln p_1}\approx
\nonumber\\
\approx-\frac{1}{p_1^2}(1+p_1^2\ln p_1).
\end{eqnarray}
Using Eq.~(\ref{eq:75}), we obtain:
\begin{eqnarray}\label{eq:126}
\lambda_2^2=\varepsilon-\left(1-\frac{p_1^2}{a^2}\right),
\nonumber\\
\varepsilon=\frac{p_2^2}{a^2}+1-\frac{p_1^2}{a^2}=1+\frac{p_2^2-p_1^2}{a^2}.
\end{eqnarray}
Therefore,
\begin{widetext}
\begin{eqnarray}\label{eq:127}
D\approx\left(1-\frac{p_1^2}{a^2}\right)\left(\frac{1}{p_1^2}-\frac{1}{p_2^2}\right)^2-\left(\frac{1}{p_1^2}+\ln p_1+f_2(p_2)\right)
\left(\frac{1}{p_1^2}+\ln p_1+f_2(p_2)\left(1+\frac{p_2^2-p_1^2}{a^2}\right)\right)=
\nonumber\\
=\left(1-\frac{p_1^2}{a^2}\right)\left(\frac{1}{p_1^4}-\frac{2}{p_1^2 p_2^2}+\frac{1}{p_2^4}\right)-\left(\frac{1}{p_1^2}+\ln p_1+f_2(p_2)\right)^2-\left(\frac{1}{p_1^2}+\ln p_1+f_2(p_2)\right)f_2(p_2)\frac{p_2^2-p_1^2}{a^2}=
\nonumber\\
=\frac{1}{p_1^2}\left(-\frac{2}{p_2^2}-\frac{1}{a^2}-2 f_3-f_2\frac{p_2^2}{a^2}\right)+\left(\frac{1}{p_2^4}+\frac{2}{a^2 p_2^2}-f_3^2+\frac{f_2}{a^2}-f_3 f_2\frac{p_2^2}{a^2}\right)+p_1^2\left(-\frac{1}{a^2 p_2^4}+f_2 f_3\frac{1}{a^2}\right),
\end{eqnarray}
\end{widetext}
where
\begin{equation}\label{eq:128}
f_3=\ln p_1+f_2(p_2).
\end{equation}
We further assume that the absolute value of the term with $\frac{1}{p_1^2}$ in the last line of Eq.~(\ref{eq:127}) is greater than or of the same order of magnitude as the absolute values of the two other terms. As $D$ enters the denominator of Eq.~(\ref{eq:120}), we may now neglect the two lesser terms: for any domain of parameters providing high absorption efficiency that we may find in this way, the assumption will ensure high absorption efficiency after we take into account the two neglected terms. On the whole, our assumptions may be loosely defined in the following way: $p_1$ (which is proportional to the ratio of the radii of the cylinder and the beam waist) is "much less than any other parameter of the problem".
Thus, we obtain
\begin{equation}\label{eq:129}
D p_1^2\approx-\frac{2}{p_2^2}-\frac{1}{a^2}-2 \ln p_1-2 f_2-f_2\frac{p_2^2}{a^2},
\end{equation}
and
\begin{eqnarray}\label{eq:130}
\frac{W^a}{W}=\frac{-16\textrm{Im}\left(\frac{1}{p_2^2}+\frac{\varepsilon+1}{2}\frac{J'}{p_2 J}\right)}{\left|-\frac{2}{p_2^2}-\frac{1}{a^2}-2 \ln p_1-2 f_2-f_2\frac{p_2^2}{a^2}\right|^2}=
\nonumber\\
=\frac{-16\textrm{Im}\left(\frac{1}{p_2^2}+\left(1+\frac{p_2^2-p_1^2}{2 a^2}\right)\frac{J'}{p_2 J}\right)}{\left|-\frac{2}{p_2^2}-\frac{1}{a^2}-2 \ln p_1-2 f_2-f_2\frac{p_2^2}{a^2}\right|^2}\approx
\nonumber\\
\approx\frac{-16\textrm{Im}\left(\frac{1}{p_2^2}+f_2\left(1+\frac{p_2^2}{2 a^2}\right)\right)}{\left|-\frac{2}{p_2^2}-\frac{1}{a^2}-2 \ln p_1-2 f_2\left(1+\frac{p_2^2}{2 a^2}\right)\right|^2}=
\nonumber\\
=\frac{-16\textrm{Im}(f_4)}{\left|-2 f_4-\frac{1}{a^2}-2 \ln p_1\right|^2}=\frac{-4\textrm{Im}(f_4)}{\left|-f_4-\frac{1}{2 a^2}-\ln p_1\right|^2}=
\nonumber\\
=\frac{4\textrm{Im}(f_5)}{\left|f_5-\ln p_1\right|^2}=\frac{4\textrm{Im}(f_6)}{\left|f_6\right|^2}=\frac{4}{2 i}\frac{f_6-f_6^*}{|f_6|^2}=
\nonumber\\
=\frac{2}{i}\left(\frac{1}{f_6^*}-\frac{1}{f_6}\right)=4\textrm{Im}\left(\frac{1}{f_6^*}\right)=
\nonumber\\
=-4\textrm{Im}\left(\frac{1}{f_6}\right)=
\textrm{Im}(f_7),\hspace{10pt}
\end{eqnarray}
where
\begin{eqnarray}\label{eq:131}
f_4=\frac{1}{p_2^2}+f_2\left(1+\frac{p_2^2}{2 a^2}\right), f_5=-f_4-\frac{1}{2 a^2},
\nonumber\\
f_6=f_5-\ln p_1, f_7=-\frac{4}{f_6}=-\frac{4}{f_5-\ln p_1.}
\end{eqnarray}
Therefore,
\begin{eqnarray}\label{eq:132}
f_5=-\frac{1}{p_2^2}-f_2\left(1+\frac{p_2^2}{2 a^2}\right)-\frac{1}{2 a^2}=
\nonumber\\
=-f_2\left(1+\frac{p_2^2}{2 a^2}\right)-\frac{1}{p_2^2}\left(1+\frac{p_2^2}{2 a^2}\right)=
\nonumber\\
=-\left(f_2+\frac{1}{p_2^2}\right)\left(1+\frac{p_2^2}{2 a^2}\right)=
\nonumber\\
=-(1+f_2 p_2^2)\left(\frac{1}{p_2^2}+\frac{1}{2 a^2}\right).
\end{eqnarray}
On the other hand,
\begin{eqnarray}\label{eq:133}
1+f_2 p_2^2=1+p_2\frac{J'_1(p_2)}{J_1(p_2)}=p_2\frac{\frac{J_1(p_2)}{p_2}+J'_1(p_2)}{J_1(p_2)}=
\nonumber\\
=p_2\frac{J_0(p_2)}{J_1(p_2)}.
\end{eqnarray}
Thus, from Eqs.~(\ref{eq:130}), (\ref{eq:131}), (\ref{eq:131}), we obtain the asymptotic formula for absorption efficiency:
\begin{eqnarray}\label{eq:134}
\eta=\frac{W^a}{W}=\textrm{Im} \frac{4}{p_2\frac{J_0(p_2)}{J_1(p_2)}\left(\frac{1}{p_2^2}+\frac{1}{2 a^2}\right)+\ln p_1}.
\end{eqnarray}
We see that, all other parameters being equal, the absorption efficiency depends very weakly on $p_1$, which is proportional to the ratio of the radii of the cylinder and the beam waist.
\section{\label{sec:level15}Efficient heating in the transverse geometry}
Let us now consider the transverse geometry, where a cylindrical electromagnetic wave converges on a thin conducting cylinder. The general solution of Section~\ref{sec:level13} may be used in this case with $\beta=0$, $n=0$, $\gamma=0$, but this relatively simple case is worth direct consideration, as it is easier to verify, and it is more natural to choose the incident TM wave using the Hankel function of the second kind rather than the Bessel function, as the incident wave propagates in the radial direction (towards the axis) rather than along the axis. Converging TE waves are not considered here, as they do not provide efficient heating in the relevant parameter domain.
\begin{eqnarray}\label{eq:573}
\Pi^{inc}=H^{(2)}_0(\rho),
\nonumber\\
\bm{H}^{inc}=\left\{0,i
H^{(2)'}_0(\rho),0\right\},
\nonumber\\
\bm{E}^{inc}=\left\{0,0,H^{(2)}_0(\rho)\right\}.
\end{eqnarray}
Here $H^{(2)}_n$ is the Hankel function of the second kind. We assume here that $\mu_1=\mu_2=1$, $\varepsilon_1=1$, $\varepsilon_2=\varepsilon$,$k_0=1$. Again, the actual electric and magnetic fields are equal to the real parts of the products of the above expressions by the time-dependent factor $\exp(-i\omega t)$.
A diverging reflected cylindrical wave propagates from the cylinder outwards:
\begin{eqnarray}\label{eq:574}
\Pi^{r\!fl}=a_1H^{(1)}_0(\rho),
\nonumber\\
\bm{H}^{r\!fl}=a_1\left\{0,i
H^{(1)'}_0(\rho),0\right\},
\nonumber\\
\bm{E}^{r\!fl}=a_1\left\{0,0,H^{(1)}_0(\rho)\right\}.
\end{eqnarray}
Inside the cylinder, the refracted wave has the following form:
\begin{eqnarray}\label{eq:575}
\Pi^{r\!fr}=a_2J_0(\sqrt{\varepsilon}\rho),
\nonumber\\
\bm{H}^{r\!fr}=a_2\left\{0,i
\sqrt{\varepsilon}J^{'}_0(\sqrt{\varepsilon}\rho),0\right\},
\nonumber\\
\bm{E}^{r\!fr}=a_2\left\{0,0,J_0(\sqrt{\varepsilon}\rho)\right\}.
\end{eqnarray}
The tangential components of the electric and magnetic fields must be continuous at the surface of the cylinder,
so
\begin{eqnarray}\label{eq:576}
H^{inc}_{\varphi}+H^{r\!fl}_{\varphi}=H^{r\!fr}_{\varphi},
\nonumber\\
E^{inc}_z+E^{r\!fl}_z=E^{r\!fr}_z,
\nonumber\\
iH^{(2)'}_0(a)+ia_1H^{(1)'}_0(a)=i a_2\sqrt{\varepsilon}J^{'}_0(\sqrt{\varepsilon}a),
\nonumber\\
H^{(2)}_0(a)+a_1H^{(1)}_0(a)=a_2J_0(\sqrt{\varepsilon}a).
\end{eqnarray}
Hence, if we denote $\sqrt{\varepsilon} a=p$, then
\begin{equation}\label{eq:577}
b=\frac{H^{(2)'}_0(a)H^{(1)}_0(a)-H^{(2)}_0(a)H^{(1)'}_0(a)}{\sqrt{\varepsilon}J^{'}_0(p)H^{(1)}_0(a)-J_0(p)H^{(1)'}_0(a)}.
\end{equation}
Using the formula for the Wronskian of Hankel functions, we obtain:
\begin{eqnarray}\label{eq:578}
H^{(2)'}_0(a)H^{(1)}_0(a)-H^{(2)}_0(a)H^{(1)'}_0(a)=
\nonumber\\
=W\{H^{(1)}_0(a),H^{(2)}_0(a)\}=-\frac{4i}{\pi a},
\end{eqnarray}
so
\begin{equation}\label{eq:577a}
a_2=-\frac{4i}{\pi a(\sqrt{\varepsilon}J^{'}_0(p)H^{(1)}_0(a)-J_0(p)H^{(1)'}_0(a))}.
\end{equation}
The Poynting vector averaged over a period can be expressed in terms of complex amplitudes of electrical and magnetic fields as $\frac{1}{2}\frac{c}{4\pi} \textrm{Re}[\bm{E}\times\bm{H}^{*}]$ (the asterisk denotes complex conjugation). The heating efficiency $\eta$ (the share of the power in the incident wave that is absorbed in the cylinder) equals the ratio of the radial components of energy flows in the refracted wave (Eq.~(\ref{eq:575})) and in the  incident wave (Eq.~(\ref{eq:573})) at the surface of the cylinder. These components do not depend on the azimuthal angle $\varphi$, so
\begin{eqnarray}\label{eq:578a}
\eta=\frac{\textrm{Re}[\bm{E}^{r\!fr}\times\bm{H}^{r\!fr*}]_{\rho}}{\textrm{Re}[\bm{E}^{inc}\times\bm{H}^{inc*}]_{\rho}}=\frac{\textrm{Re}(-E^{r\!fr}_z H^{r\!fr*}_{\varphi})}{\textrm{Re}(-E^{inc}_z H^{inc*}_{\varphi})}=
\nonumber\\
=\frac{E^{r\!fr}_z H^{r\!fr*}_{\varphi}+E^{r\!fr*}_z H^{r\!fr}_{\varphi}}{E^{inc}_z H^{inc*}_{\varphi}+E^{inc*}_z H^{inc}_{\varphi}}=
\nonumber\\
=\frac{|a_2^2|(J_0(p)(-i(\sqrt{\varepsilon})^*)J^{'*}_0(p)+J^*_0(p)i\sqrt{\varepsilon}J^{'}_0(p))}{H^{(2)}_{0}(a)(-i)H^{(1)'}_0(a)+H^{(1)}_0(a)iH^{(2)'}_0(a)}=
\nonumber\\
=\frac{|a_2^2|i(J^*_0(p)\sqrt{\varepsilon}J^{'}_0(p)-J_0(p)(\sqrt{\varepsilon})^*J^{'*}_0(p))}{i\left(-\frac{4i}{\pi a}\right)}.\hspace{10pt}
\end{eqnarray}
We are interested in the parameter domain where the conductivity of the cylinder is high and its radius is much less than the wavelength. So let us assume now that $\varepsilon=\varepsilon'+i\varepsilon''$, where $\varepsilon'$ and $\varepsilon''$ are real, $\varepsilon''\gg|\varepsilon'|$, and
\begin{eqnarray}\label{eq:579}
a\ll 1, |p|\ll 1, |\ln a|\gg 1.
\end{eqnarray}
Then $J_0(p)\approx 1$,  $J^{'}_0(p)\approx-\frac{p}{2}$, $H^{(1)}_0(a)\approx i\frac{2}{\pi}\ln a$, $H^{(1)'}_0(a)\approx i\frac{2}{\pi a}$, so Eq.~(\ref{eq:577}) becomes:
\begin{eqnarray}\label{eq:580}
a_2\approx-\frac{4i}{\pi a(\sqrt{\varepsilon}(-\frac{p}{2})i\frac{2}{\pi}\ln a-i\frac{2}{\pi a})}=
\nonumber\\
=\frac{4}{\varepsilon a^{2}\ln a+2},
\end{eqnarray}
so
\begin{eqnarray}\label{eq:581}
|a_2^2|=\frac{16}{|\varepsilon a^2\ln a+2|^2}\approx\frac{16}{(\varepsilon'')^{2}a^{4}\ln^{2}a+4}.
\end{eqnarray}
On the other hand,
\begin{eqnarray}\label{eq:582}
J^*_0(p)\sqrt{\varepsilon}J^{'}_0(p)-J_0(p)(\sqrt{\varepsilon})^*J^{'*}_0(p)\approx
\nonumber\\
\approx \sqrt{\varepsilon}(-\frac{p}{2})-(\sqrt{\varepsilon})^{*}(-\frac{p^{*}}{2})=(\varepsilon-\varepsilon^{*})(-\frac{a}{2})=
\nonumber\\
=-i\varepsilon''a.
\end{eqnarray}
Thus,
\begin{eqnarray}\label{eq:584}
\eta=\frac{\frac{16}{(\varepsilon'')^{2}a^{4}\ln^{2}a+4}\varepsilon'' a}{\frac{4}{\pi a}}=\frac{4\pi\varepsilon'' a^2}{(\varepsilon'')^{2}a^{4}\ln^{2}a+4}.
\end{eqnarray}
If
\begin{eqnarray}\label{eq:585}
\varepsilon''a^{2}|\ln a|\sim 2,
\end{eqnarray}
where $\sim$ means "is of the same order of magnitude as", then
\begin{eqnarray}\label{eq:586}
\eta=\frac{\pi}{|\ln a|}\frac{2}{\frac{\varepsilon''a^{2}|\ln a|}{2}+\frac{2}{\varepsilon''a^{2}|\ln a|}}\sim \frac{\pi}{|\ln a|}.
\end{eqnarray}
It is evident that the heating efficiency decreases extremely slowly when the radius of the cylinder decreases by many orders of magnitude, if in the same time the conductivity of the cylinder increases to satisfy the optimality condition (Eq.~(\ref{eq:585})). In this case the efficiency typically equals $20\div40\%$. Moreover, the condition of high efficiency is not resonant, as the width of the maximum corresponds to variation of conductivity by an order of magnitude (at the level of over 50\% of the maximum).

It should also be noted that Eq.~(\ref{eq:585}) is compatible with the approximations (Eq.~(\ref{eq:579})).
\maketitle
\section{\label{sec:level16}Conclusions}
A rigorous analysis of a diffraction problem demonstrates that it is possible to achieve efficient heating of cylindrical targets by electromagnetic beams with transverse dimensions that are several orders of magnitude greater than those of the cylinder. The relevant conditions are not too rigid.

This is the first part of the work. It contains the following results:

Exact solutions of the free Maxwell equations are described that are approximated by Gaussian beams with great accuracy when the beam waist radius is much greater than the wavelength, and their asymptotic properties are proved. It is also proven that these solutions are normalizable. These results may be useful for many applications.

The qualitative, semiquantitative, and quantitative approaches are developed for calculation of efficiency of heating of cylindrical targets with broad electromagnetic beams. The quantitative approach includes rigorous solution of the Maxwell equations.

Asymptotic formulae for heating efficiency in the longitudinal and transverse geometry are derived.

\begin{acknowledgments}
The author is very grateful to Dr. A.V. Gavrilin for valuable remarks and help.
\end{acknowledgments}
\bibliography{dfcdb1}

\end{document}